\documentclass[aps,prd,onecolumn,eqsecnum,amsmath,nofootinbib,preprintnumbers]{revtex4}%

\usepackage{color,graphicx,float,subfigure,xcolor}
\usepackage{amsfonts,amssymb,theorem,mathrsfs,times}
\usepackage{bm}
\usepackage{mathtools}
\usepackage{amsfonts,amssymb,theorem,mathrsfs}
\usepackage{dsfont}
\usepackage{setspace}
\usepackage{amsmath}  
\usepackage{soul, color, xcolor}
\usepackage[colorlinks,linkcolor=blue,anchorcolor=blue,citecolor=blue]{hyperref}   
\usepackage{ulem}   
\usepackage[english]{babel}

{\theorembodyfont{\upshape}
	}
{\theorembodyfont{\upshape}
	}
{\theorembodyfont{\upshape}
	}
{\theorembodyfont{\upshape}
	}
{\theorembodyfont{\upshape}
	}
{\theorembodyfont{\upshape}
	}
\addtocounter{MaxMatrixCols}{10}
\newcommand{\dalm}{\kern1pt\vbox{\hrule height 0.9pt\hbox{\vrule width
			0.9pt\hskip 2.5pt\vbox{\vskip 5.5pt}\hskip 3pt\vrule width
			0.3pt}\hrule height 0.3pt}\kern1pt}

\begin{document}
\thispagestyle{empty}
\preprint{\hfill {\small {ICTS-USTC/PCFT-24-03}}}
\title{The black hole shadow of quantum Oppenheimer-Snyder–de Sitter spacetime }
%
\author{ Shu Luo$^{a}$\footnote{e-mail
			address: ls040629@mail.ustc.edu.cn}}
\author{Cheng-Hao Li$^{b}$\footnote{e-mail address: chris0310@mail.ustc.edu.cn}}

\affiliation{${}^a$School of  Physics ,
 University of Science and Technology of China, Hefei, Anhui 230026,
	China}

\affiliation{${}^a$School of  Gifted Young ,
 University of Science and Technology of China, Hefei, Anhui 230026,
	China}

\date{\today}
	
\begin{abstract}
In this study, we investigate the black hole shadow of an exact black hole solution of the loop quantum gravity (LQG) theory. We discuss some of its optical characteristics after generalizing it to the rotational case, including null geodesics and black hole shadow. From these we can compare the impact of different theories on the most deeply understood properties of the black hole and get a new way to test the accuracy of the modified gravity theory. 
\end{abstract}

\maketitle

\section{Introduction}
Recently, with the detection of gravitational waves (GW) from binary compact stars~\cite{LIGOScientific:2016aoc}, together with the Event Horizon Telescope (EHT) revealing the optical characteristics of M87 center supermassive black hole~\cite{Akiyama_2019},  black holes, especially astronomical realistic black holes, can serve  as a lab to test various theoretical predictions,  including quantum gravity and other modified gravity theories through Multi Messenger Astronomy. These modifications mainly consists of two types: adding new elements to the right hand of the Einstein equation, i.e., supposing new source of energy-momentum tensor to tangle with gravitation, including dark matter and dark energy~\cite{Salucci_2000,Burkert_1995}; modifying the left side of the equation by the new gravitational degrees of freedom in addition to the metric tensor. In many cases, the modification of GR can be described by a scalar-tensor theory of gravitation~\cite{CLIFTON20121,Berti_2015}. Moreover, to solve the difficult task when dealing with singularities that are unavoidable in GR, a new candidate for unified theory, Loop quantum Gravity (LQG) provides us with several new features.

Black hole shadow of different kinds of black holes is also an intensively investigated area of black hole characteristics and possible tests of modified gravity theories. Theoretically, the shadow cast by the black hole horizon is studied as null geodesics and the existence of
a photon sphere is expected. The incoming photons are trapped in an unstable circular orbit. Occasionally the photons are perturbed and diverted towards the observer. For most common axisymmetric black hole the two celestial coordinates $\alpha$ and $\beta$  are the observable quantities. After the pioneering work of Synge~\cite{10.1093/mnras/131.3.463}, both Kerr and Kerr-Newman black hole's shadow have been investigated. Apart from vacuum solution, Kerr black hole in matter field are also intensively studied, including in perfect-fluid dark matter~\cite{PhysRevD.99.044015} and Burkert  halo~\cite{PhysRevD.100.044012}. Also there are a lot of studies focused on the test of different gravity models through image of black hole shadow:
Einstein-Gauss-Bonnet BH~\cite{CUNHA2017373}; Konoplya-Zhidenko BH~\cite{Wang_2017}; traversable wormhole~\cite{PhysRevD.98.024044}; Kerr black hole in plasma~\cite{PhysRevD.95.104003}, RN-dS black hole~\cite{Cao:2024kht} and so on.
Recently the images of many compact objects were studied by the ray-tracing method proposed in~\cite{PhysRevD.100.024018}, such as Kazakov-Solodukhin black hole, the regular black hole with de Sitter core, the quantum-corrected black hole, the wormholes and so on. All of these works have exhibited the reliability of the way to get the optical appearance of the black holes or other compact objects. 

Recently, Lewandowski\,\textit{et al.}\,have proposed a new quantum black hole model (qOS) within the framework of Loop Quantum Gravity (LQG) theory~\cite{PhysRevLett.130.101501}. The exterior spacetime of this model is a suitably deformed Schwarzschild black hole. This quantum corrected black hole exhibits the same asymptotic behavior as the Schwarzschild black hole and it is stable against test scalar and vector fields by the analysis of QNMs~\cite{PhysRevD.108.104004}. Through the similar method, the solution describing a quantum black hole with a positive cosmological constant (qOS-dS) is obtained in~\cite{PhysRevD.109.064012}. While there have been several studies on the black hole shadow of this spacetime, in order to accord to experimental tests a generalization of the spherically symmetric solution to rotational black hole solution using the well-known Newman-Janis algorithm~\cite{Diego} is needed. 

There are a few new issues and features when one is aiming to study the image of a black hole in a de-Sitter spacetime. Firstly, for such black hole, the observers have to be located at least inside the cosmological horizon, rather than the null infinity. But the image is influenced greatly by the location of the observer while there is no specific location given privileged status. So at this time one has to abandon the definition used in conventional black hole study, the celestial coordinates, and turn to field of view, which are more frequently utilized in the study of escape cone and tightly relied on a certain observer's tetrad. Usually the image formed by the rays emitted directly is more narrow compared with that of Schwarzschild one. Secondly, unlike the asymptotically flat black holes, now there are outermost stable circular orbits (OSCO) in the spacetime. A physically reasonable thin accretion disk is distributed between innermost stable circular orbit (ISCO) and OSCO. And this leads to a significant outer edge in the image, while it provides challenges to define the image in non-static region, i.e., out of the outer boundary. Thirdly, during the propagation of light, the redshift or blueshift factor in de-Sitter is quite different from the one in asymptotically flat case, for example, it tends to infinity near the cosmological horizon. So we cannot naively neglect this factor when approaching the observer as in asymptotically flat case.

In this paper we mainly study the image received by an observer of a certain distance in the spacetime generalized to rotating case using Newman-Janis algorithm from quantum Oppenheimer-Snyder–de Sitter spacetime (qOS-dS). In the first part of our study we show some well-known results about the behavior of null geodesics in spherically symmetric spacetime, with this specific metric. The difference of light ray movement from that in asymptotically flat spacetime is noted. Also, a careful classification of different kinds of effective potential with different parameter range is discussed. No matter in what range, there is only one peak between the outer horizon and the the cosmological horizon.

We then generalize the solution from spherically symmetric case to rotating case using Newman-Janis algorithm. We present the common algorithm of analysing the geodesic behavior of a Kerr-type spacetime, and it is shown that most conclusions are alike. We then confirm the physically rational demand for the value of the spin parameter given two other parameters $p$ and $q$, in which not only three horizons exist, but the observer is guaranteed to receive light from equatorial plane as well.

We also discuss specific black hole shadow of qOS-dS rotating spacetime. Here particularly we choose the field of view of ZAMO observer to describe the light ray projection, which serves as the one observed by a relatively static observer around the black hole. And through similar analytical calculation methods given by~\cite{PhysRevD.100.024018}, we are able to prove the rationality of merely considering the first few times of a light ray trace passing through the equatorial plane. This is important for the simplification of later computation, and it can shorten the time needed to simulate an image. After all preparation was done we move on to simulating results, in which we are managed to discover several meaningful alterations of the image from that in Kerr spacetime, including their intensity, sizes and shapes.  

The paper is organized as follows. In Sec.\ref{sec:2}, we analyse the moving characteristics of photons as well as massive particles in spherically symmetric qOS-dS spacetime. In sec.\ref{sec:3}, we use Newman-Janis algorithm to generalize the solution to rotating black holes, then we investigate the moving curves of massive or massless particles in this spacetime. In sec.\ref{sec:4}, we introduce the fundamental techniques in applying light tracing method to study the black hole shadow. Finally in sec.\ref{sec:5}, we present the simulating results and characteristics of the black hole shadow.

\section{spherically symmetric black hole solution and its circular orbits }\label{sec:2}
In our paper, we  focus on the exterior spacetime of the quantum corrected black hole whose original metric is given by~\cite{PhysRevLett.130.101501} and whose modification with a nonzero cosmos constant is given by~\cite{PhysRevD.109.064012}:
\begin{eqnarray}\label{metric}
	\mathrm{d}s^2=-f(r)\mathrm{d}t^2+\frac{\mathrm{d}r^2}{f(r)}+r^2(\mathrm{d}\theta^2+\sin^2\theta\mathrm{d}\phi^2)\, ,
\end{eqnarray}
where the metric function $f(r)$ reads
\begin{eqnarray}\label{metric_function}
	f(r)=1-\frac{2M}{r}-\frac{\Lambda}{3}r^2+\frac{\alpha M^2}{r^4}(1+\frac{\Lambda r^3}{6M})^2\, ,
\end{eqnarray}
with the positive parameter $\alpha=16\sqrt{3}\pi\gamma^3l^2_p$ , $l_p=\sqrt{\hbar}$ denoting the Planck length, $\gamma$ being the Immirzi parameter, $\lambda$ being the cosmos constant and $M$ standing for the mass of the black hole. 

\begin{figure}[htbp]
	\centering
\includegraphics[width=0.45\textwidth]{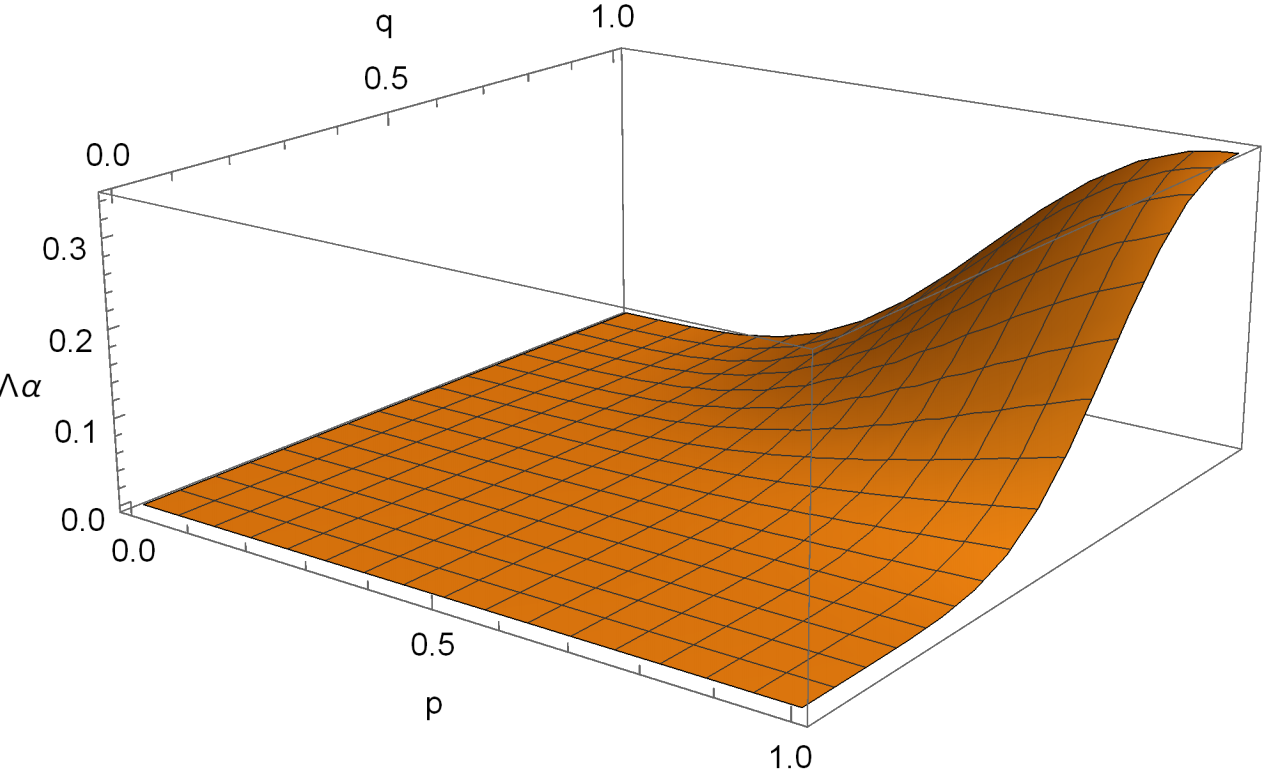}
     \includegraphics[width=0.45\textwidth]{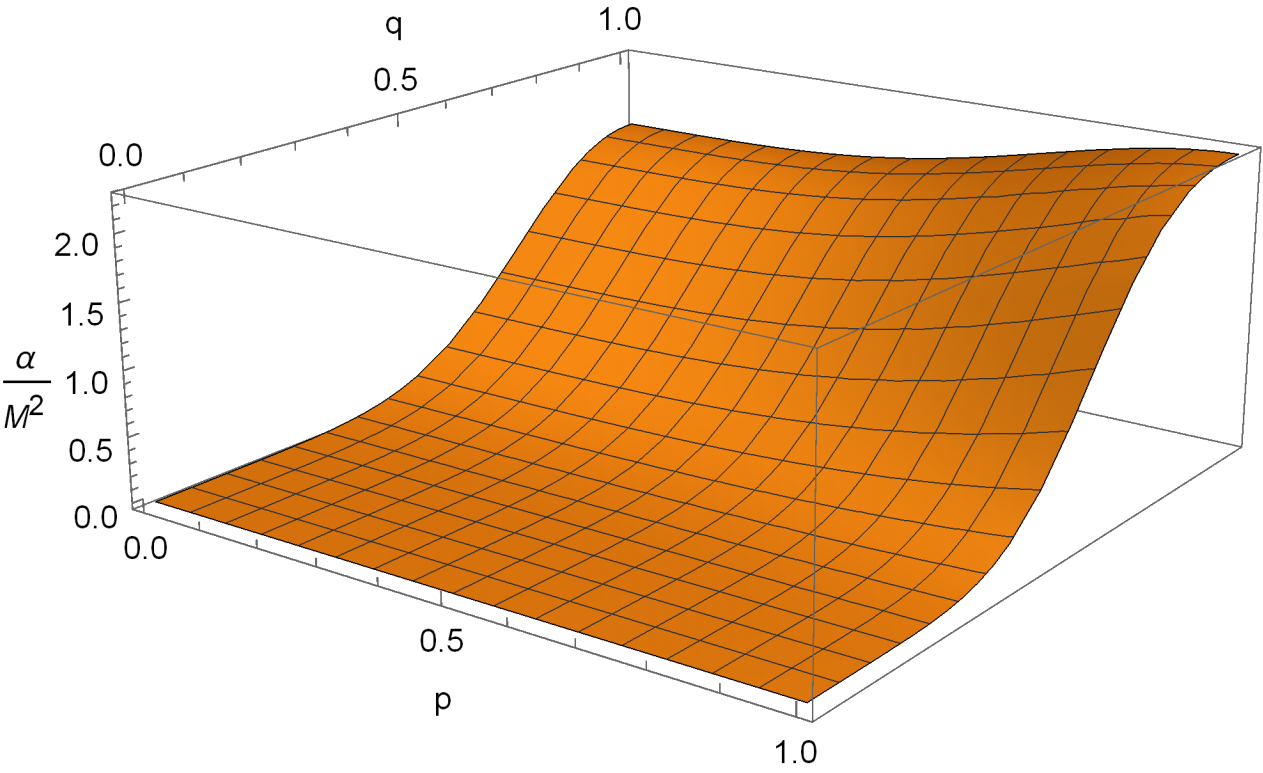}   \\ 
     \includegraphics[width=0.45\textwidth]{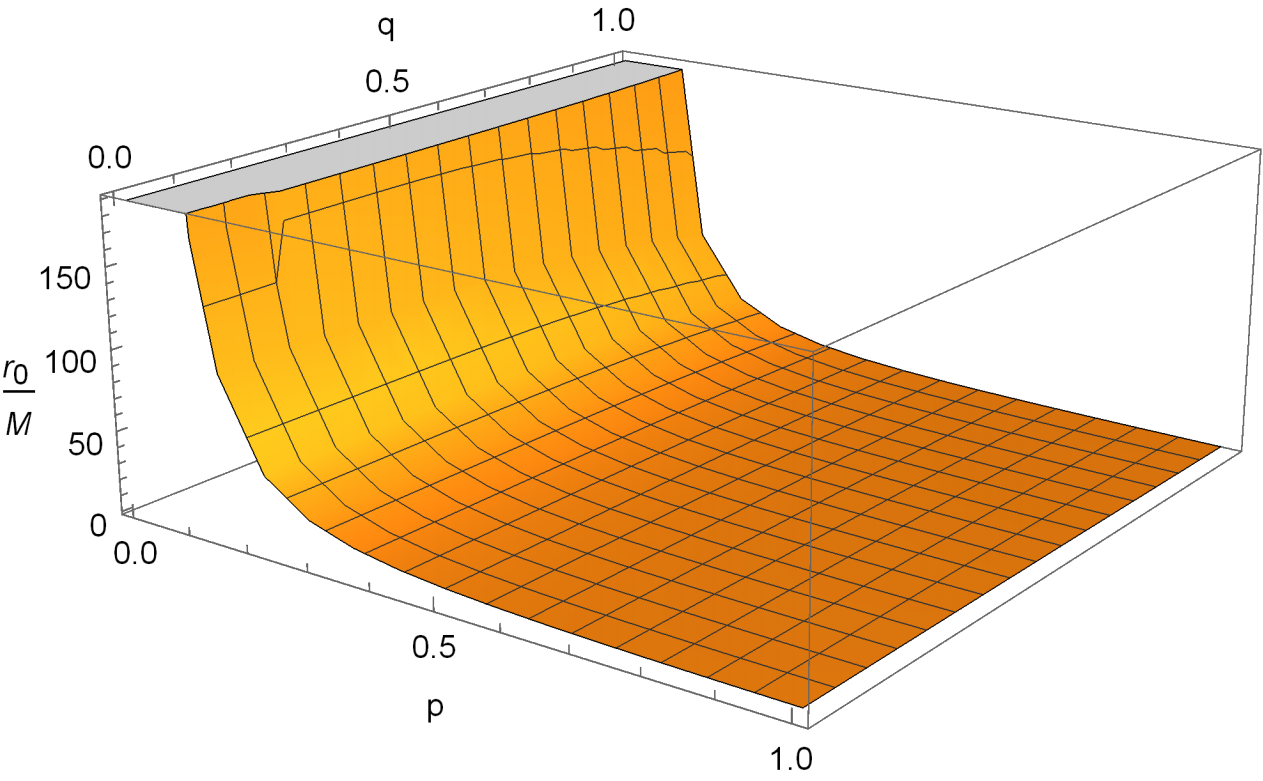}
     \includegraphics[width=0.45\textwidth]{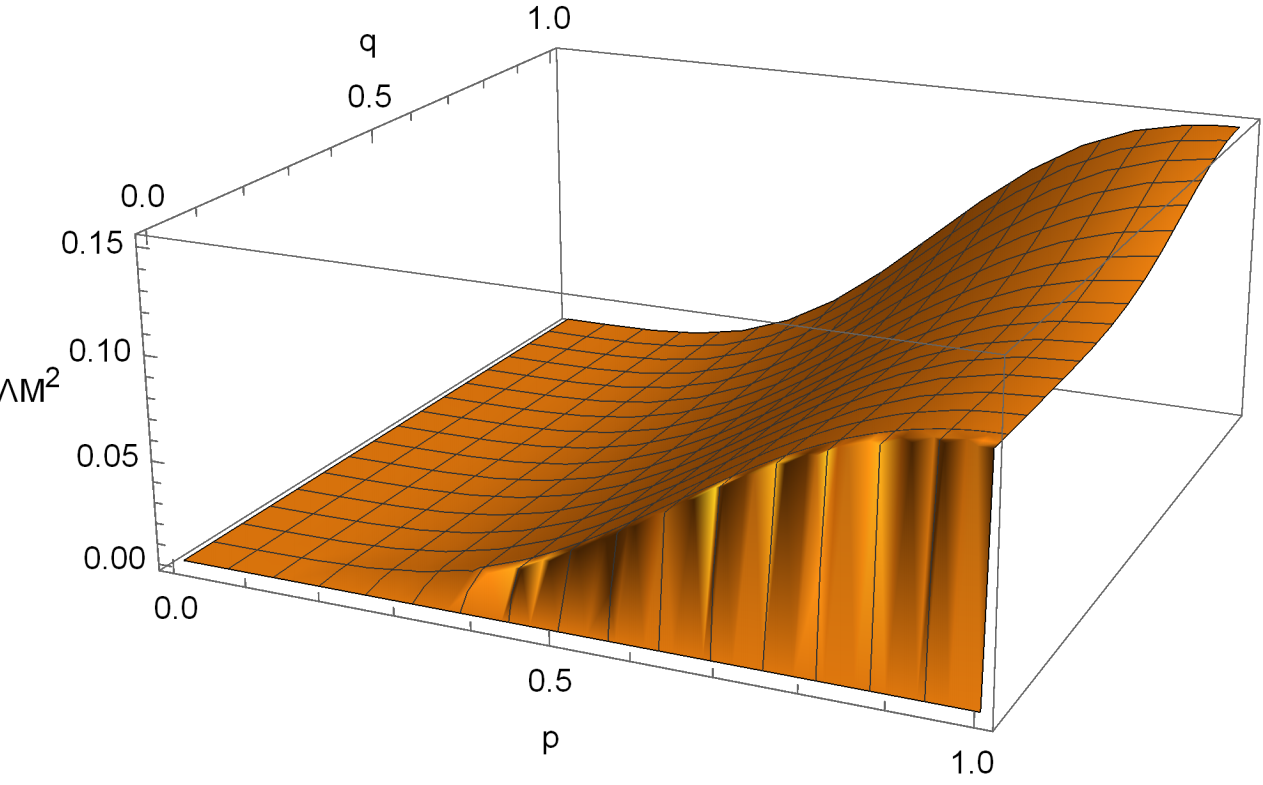}  \\
     \includegraphics[width=0.45\textwidth]{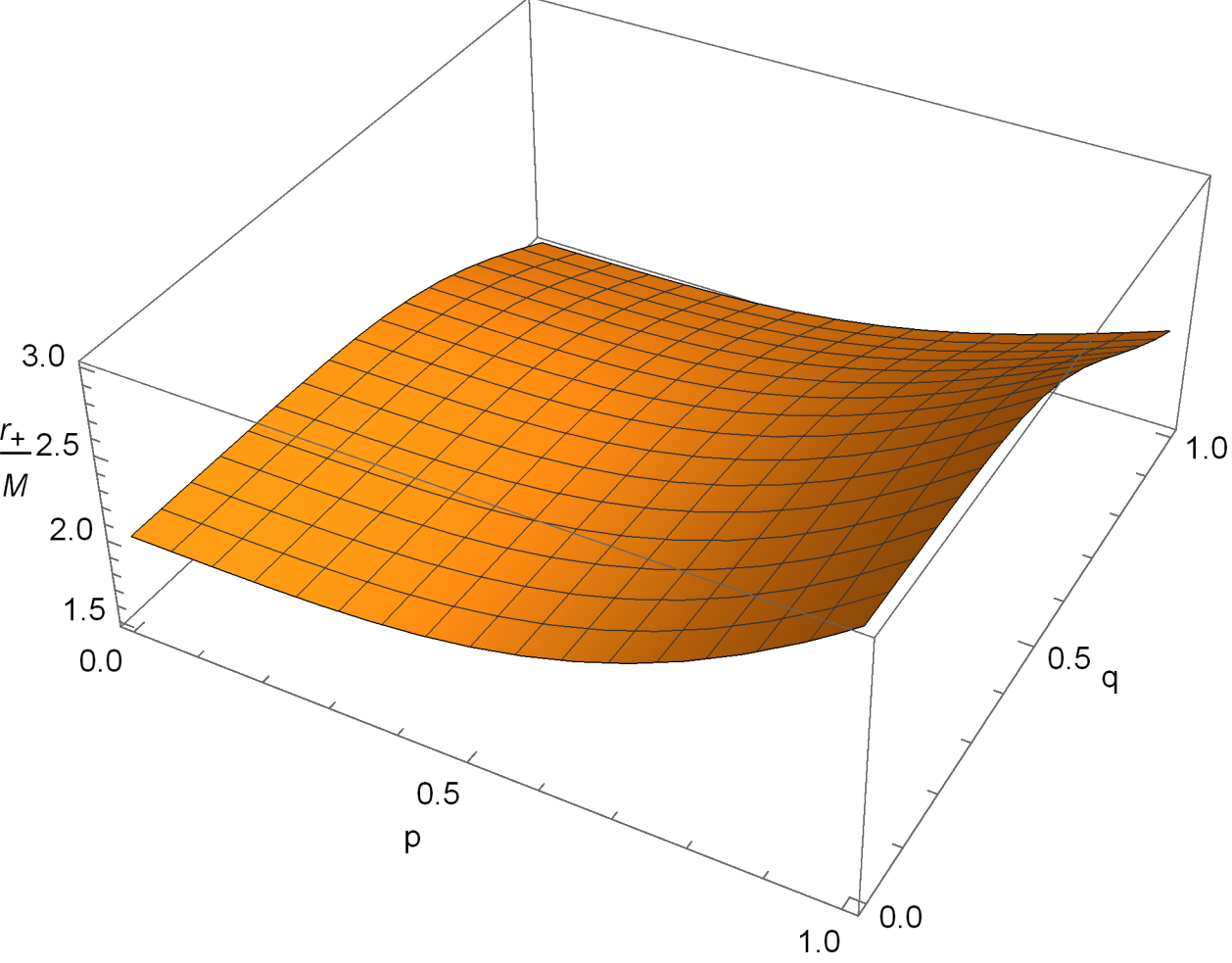} 
     \includegraphics[width=0.45\textwidth]{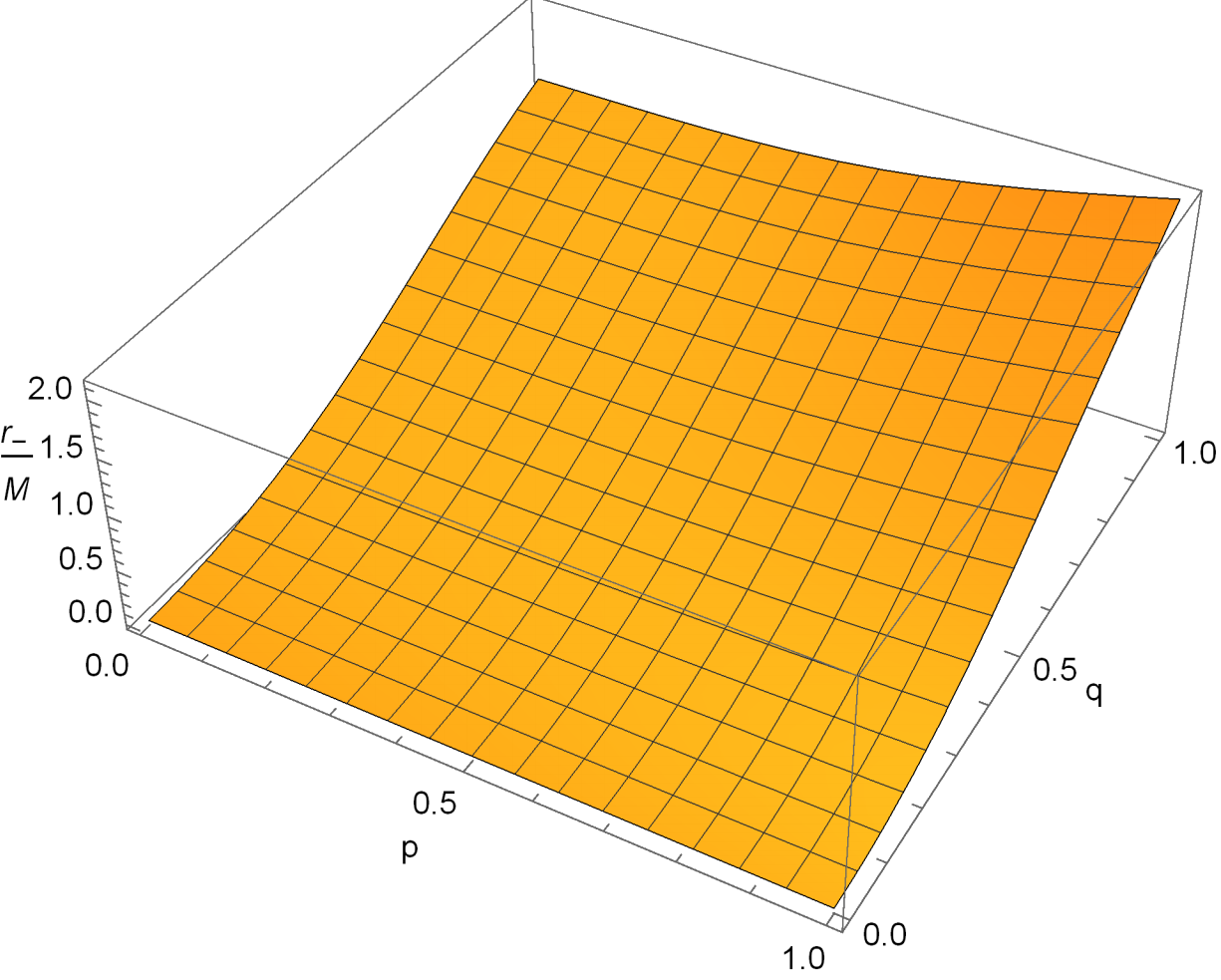} 
\caption{The relation between some given parameters and the relative ratio of the three horizon radius.  }
        \label{fig:001}
\end{figure}

The most realistic case is when there is three horizons, corresponding to three positive roots of the metric. If we set $r_{-}$, $r_{+}$ and $r_{0}$ in response to the inner, outer and cosmos horizon respectively, then~\cite{luo2024quasinormalmodespseudospectrumtime}, by setting $0<q^2=r_{-}/r_{+}<1$ and $0<p^2=r_{+}/{r_{0}}<1$,
then the metric can be rewritten as 
\begin{eqnarray}\label{oo}
    f(r)=(\frac{\alpha \Lambda^2}{36}-\frac{\Lambda}{3})r^2(1-\frac{r_{0}}{r})(1-\frac{p^2r_{0}}{r})(1-\frac{p^2q^2r_{0}}{r})(1+\frac{dr_{0}}{r}+\frac{br_{0}^2}{r^2}+\frac{cr_{0}^3}{r^3})\,,
\end{eqnarray}
where
\begin{eqnarray}
    d=1+p^2+q^2p^2\,,\quad
    b=\frac{p^2q^2(1+p^2+q^2p^2)(1+q^2+q^2p^2)}{1+q^2+q^4+p^2q^2+p^2q^4+p^4q^4}\,,\quad
     c=\frac{p^4q^4(1+p^2+q^2p^2)}{1+q^2+q^4+p^2q^2+p^2q^4+p^4q^4}\,,
\end{eqnarray}
in this way one is easy to find the relation of parameters including
\begin{eqnarray}\label{0}
   \alpha M^2r_{0}^{-4}=\tilde{a}\tilde{c}\,,\quad
   (2-\frac{\alpha \Lambda}{3})\frac{M}{r_{0}}=\tilde{b}\tilde{c}\,,\quad
   (\frac{\Lambda}{3}-\frac{\alpha \Lambda^2}{36})r_{0}^2=\tilde{c}\,,
\end{eqnarray}
where
\begin{eqnarray}\label{1}
    \tilde{a}=\frac{p^8q^6(1+p^2+q^2p^2)}{1+q^2+q^4+p^2q^2+p^2q^4+p^4q^4}\,,
\end{eqnarray}
\begin{eqnarray}\label{2}
    \tilde{b}=p^2(1+p^2)(1+q^2)\frac{1+q^4+p^4q^4+p^6q^6+p^2(q^2+q^6)}{1+q^2+q^4+p^2q^2+p^2q^4+p^4q^4}\,,
\end{eqnarray}
\begin{eqnarray}\label{3}
    \tilde{c}=\frac{1+p^2+p^4+p^2q^2+q^2p^4+p^4q^4}{(1+p^2+p^4)(1+q^2+q^4)(1+p^2q^2+p^4q^4)}\,,
\end{eqnarray}
By combing the relations in (\ref{0}) one is able to get more explicit relations between given parameters: 
\begin{eqnarray}
    \Lambda \alpha=6\Big{(}1- \sqrt{\frac{1}{1+4\tilde{a}/\tilde{b}^2}}\Big{)}\,,\quad
    \frac{\alpha}{M^2}=\frac{16\tilde{a}}{\tilde{c}^3(\tilde{b}^2+4\tilde{a})^2}\,,\quad
    \frac{r_{0}}{M}=\frac{2}{\tilde{c}}\sqrt{\frac{1}{\tilde{b}^2+4\tilde{a}}}
\end{eqnarray}
From Fig.\ref{fig:001} it is shown that the largest value of $\Lambda \alpha$ is $6-4\sqrt{2}$, if and only if $p^2=q^2=1$, or the three horizons merge, at which time there is $\alpha/M^2=9/4$ and $r_{0}/M=3\sqrt{2}/2$. 

Like many other black hole solution it takes the similar form of the RN-dS solution, where three horizons is permitted with a time-like singularity. However, just as any Cauchy horizon, the inner horizon of this black hole is extremely unstable and the actual spacetime structure inside the outer horizon might differ from the description of Eq.(\ref{metric}), which is more likely to be a null or space-like singularity, but this commonly has no influence on the metric that rules the outer space of the black hole. In this case, when considering the optical characteristics of such a black hole we only regard those light rays that go through outside the outer horizon and inside the cosmos horizon, where the observer is supposed to lie in too. 

To study the circular orbits of this solution, we follow the ordinary algorithm by writing the equation of motion 
\begin{eqnarray}\label{123}
    (\frac{\mathrm{d}r}{\mathrm{d}\tau})^2+V(r)=E^2\,,
\end{eqnarray}
and
\begin{eqnarray}\label{124}
    V(r)=(\frac{L^2}{r^2}+\delta)f(r)\,,
\end{eqnarray}
here $\delta$=0 for null geodesics and $\delta$=1 for time-like geodesics.By introducing the impact parameter $D=L/E$, Eq.(\ref{123}) and Eq.(\ref{124}) turn into 
\begin{eqnarray}\label{24}
    (\frac{\mathrm{d}u}{\mathrm{d}\theta})^2=\frac{1}{D^2}-({u}^2+\frac{\delta}{L^2})f(u),\,
\end{eqnarray}
where $u=1/r$. Just like in RN case, it's impossible for any non-radial particle to arrive at the time-like singularity, which can be illustrated by Eq.(\ref{24}) that when $u$ approaches positive infinity the right hand side will always reduce to a negative value.

\begin{figure}[htbp]
	\centering
\includegraphics[width=0.45\textwidth]{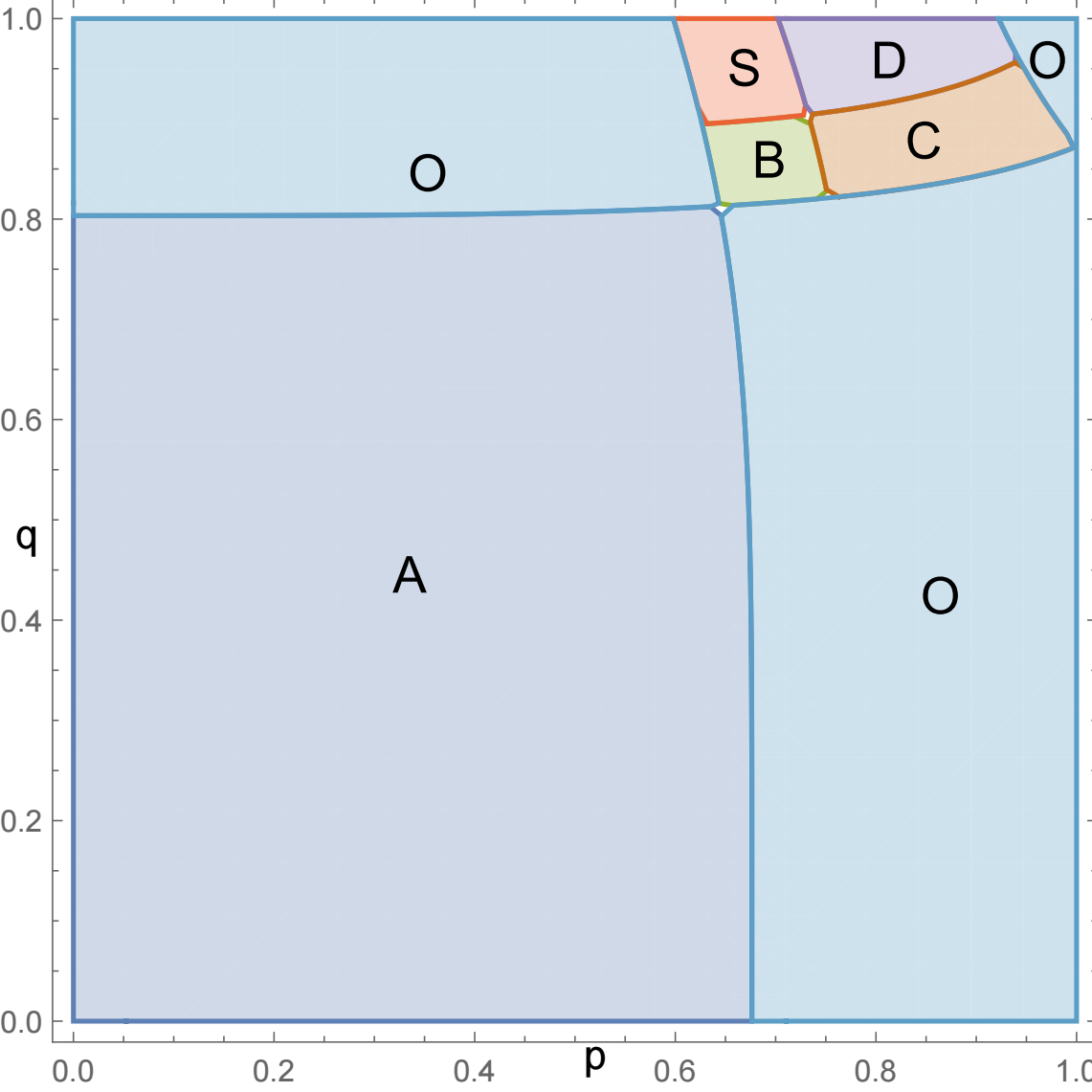}      
\caption{$p-q$ diagram and different domains for different behaviors of the effective potential. }
        \label{fig:0001}
\end{figure}

\begin{figure}[htbp]
	\centering
\includegraphics[width=0.33\textwidth]{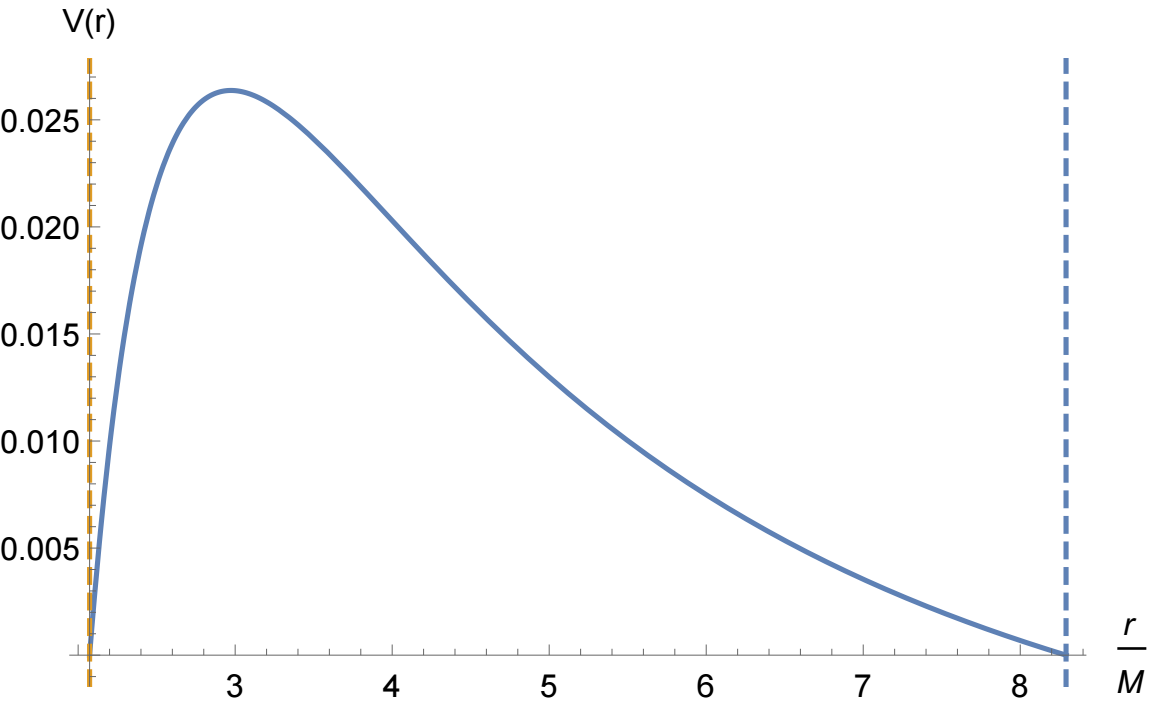}   \includegraphics[width=0.33\textwidth]{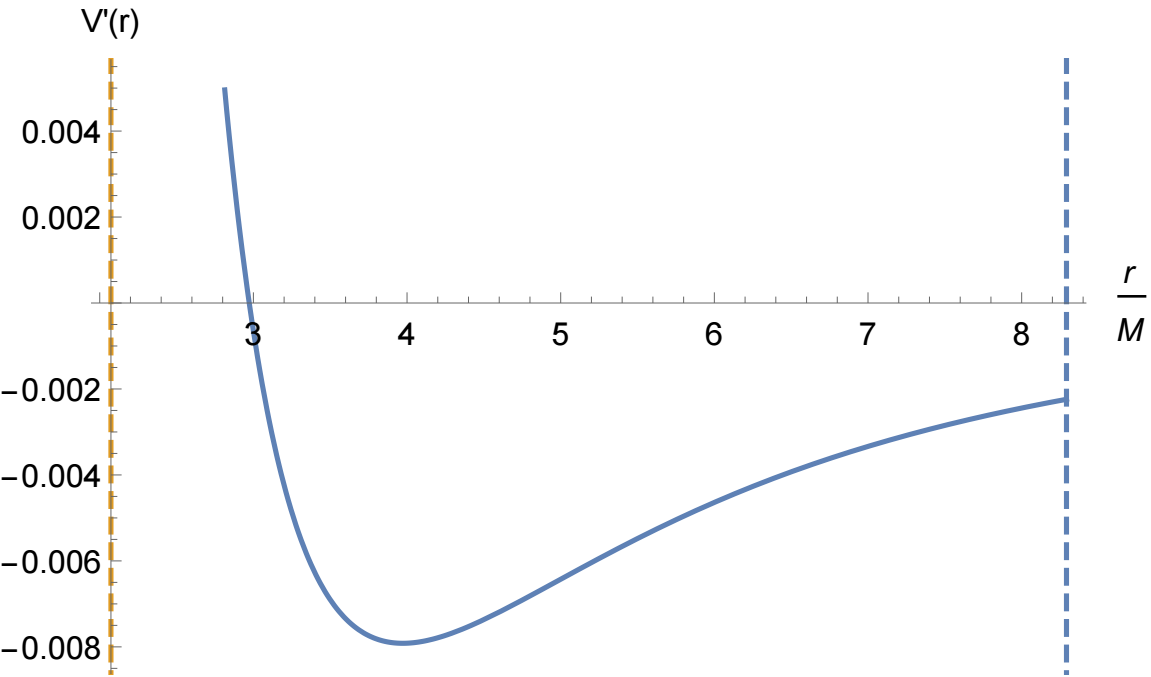}     \includegraphics[width=0.33\textwidth]{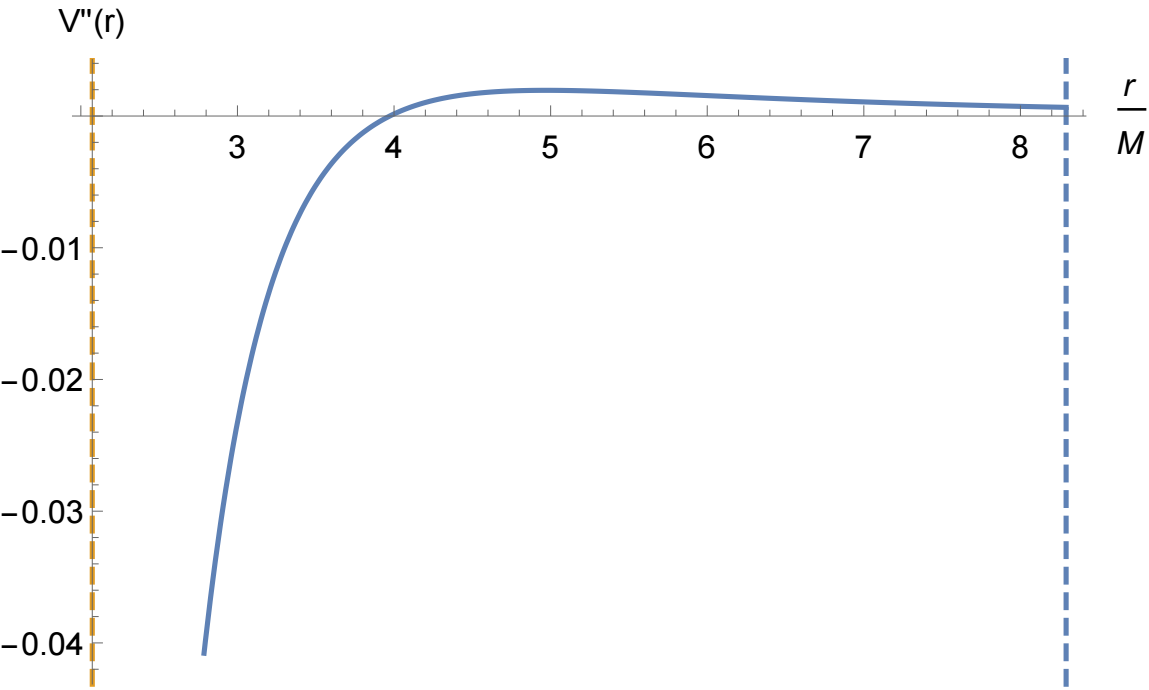}   
\caption{$V(r),V'(r),V''(r)$ as a function of $r$ from $r_{+}$ to $r_{0}$ under parameters $p=0.5,q=0.5$ in A domain.}
        \label{fig:0011}
\end{figure}

\begin{figure}[htbp]
	\centering
\includegraphics[width=0.33\textwidth]{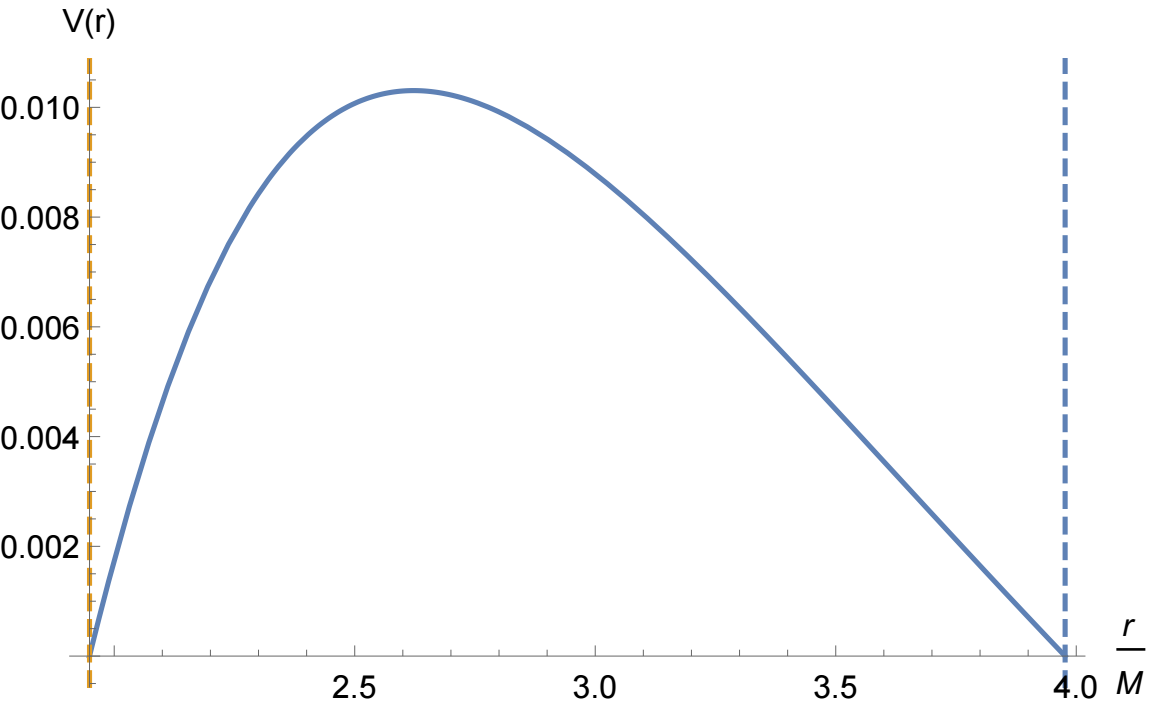}   \includegraphics[width=0.33\textwidth]{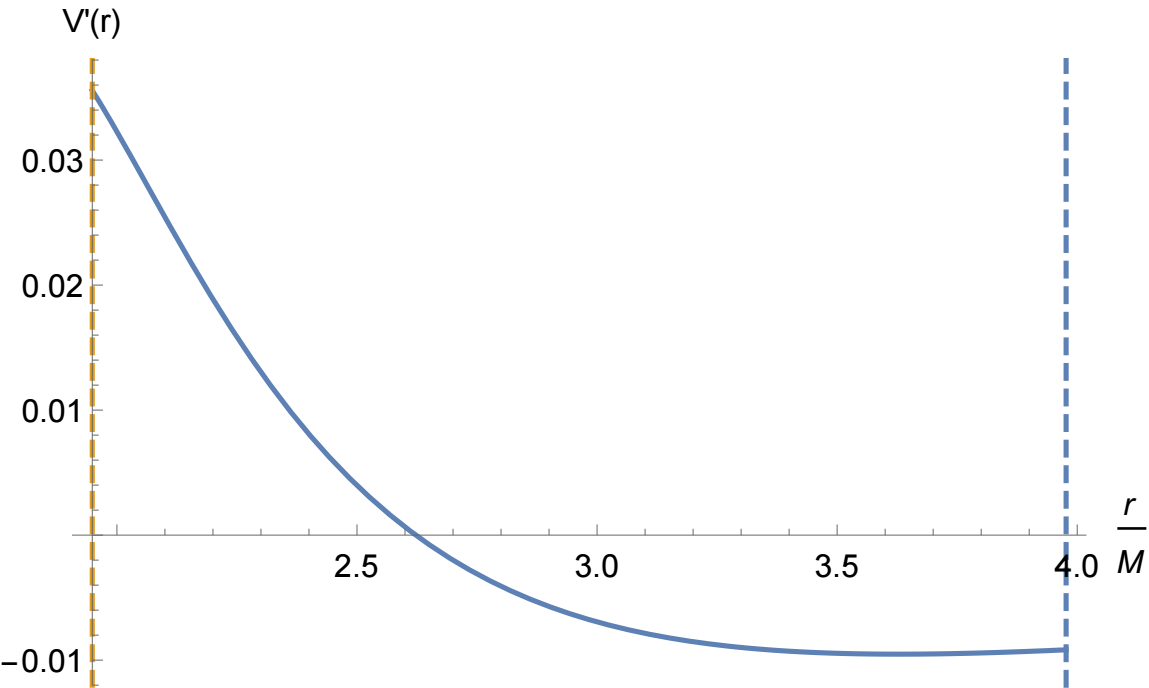}     \includegraphics[width=0.33\textwidth]{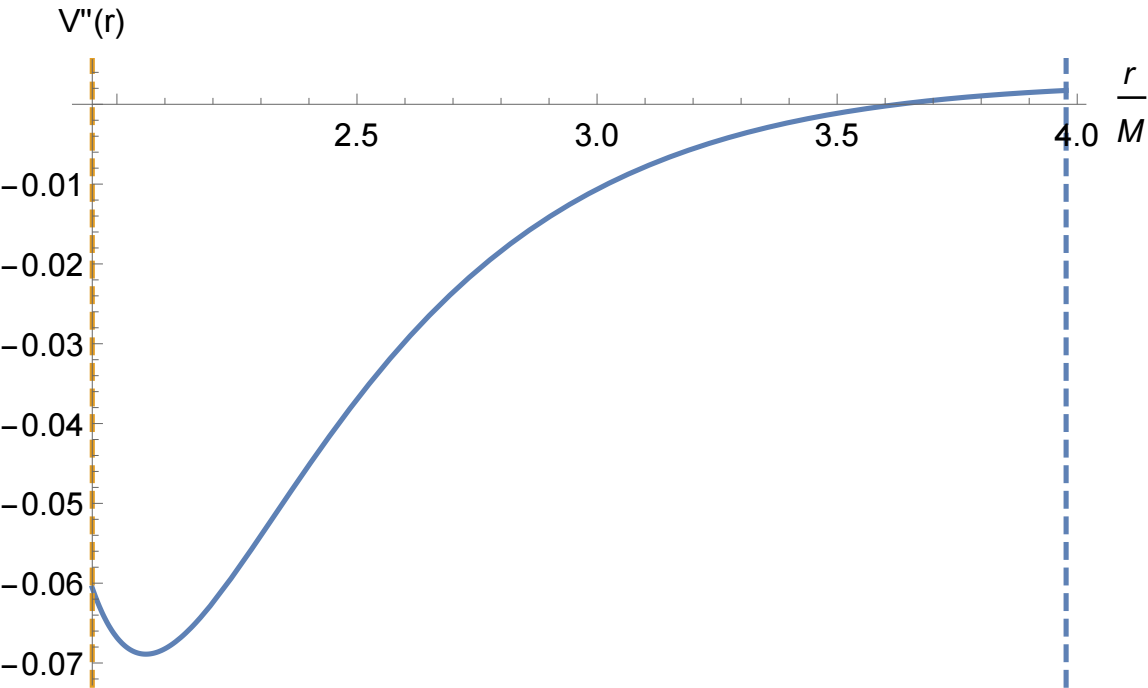}   
\caption{$V(r),V'(r),V''(r)$ as a function of $r$ from $r_{+}$ to $r_{0}$ under parameters $p=0.7,q=0.85$ in B domain.}
        \label{fig:0012}
\end{figure}

\begin{figure}[htbp]
	\centering
\includegraphics[width=0.33\textwidth]{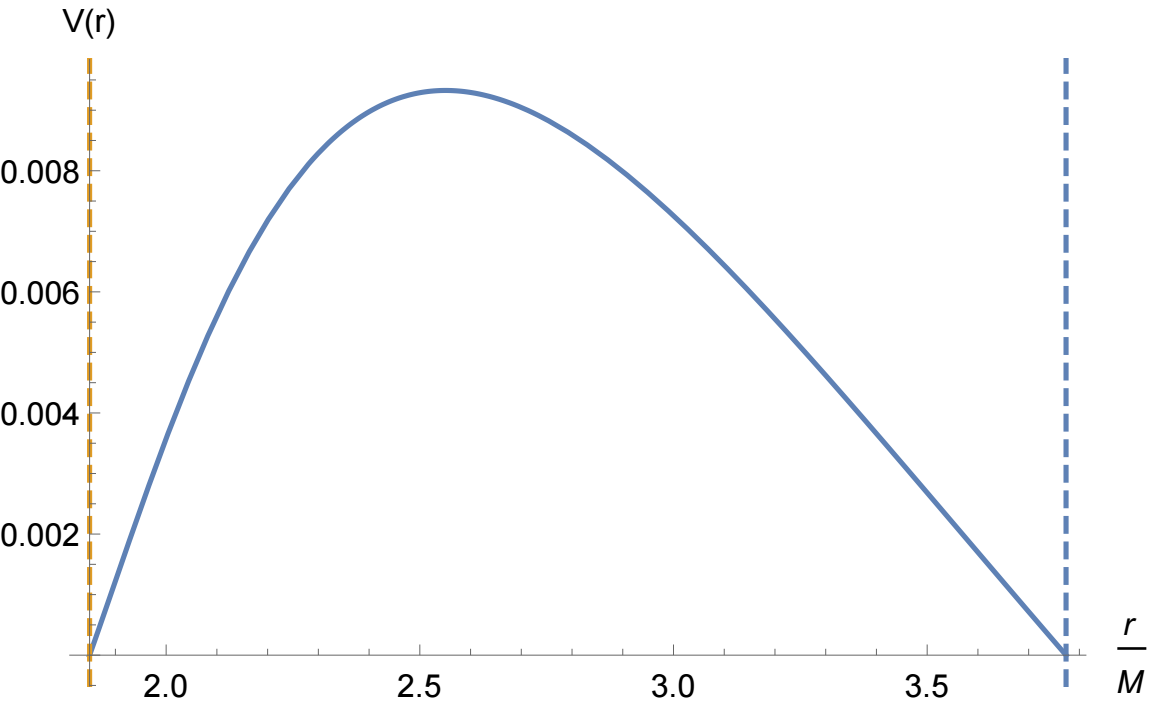}   \includegraphics[width=0.33\textwidth]{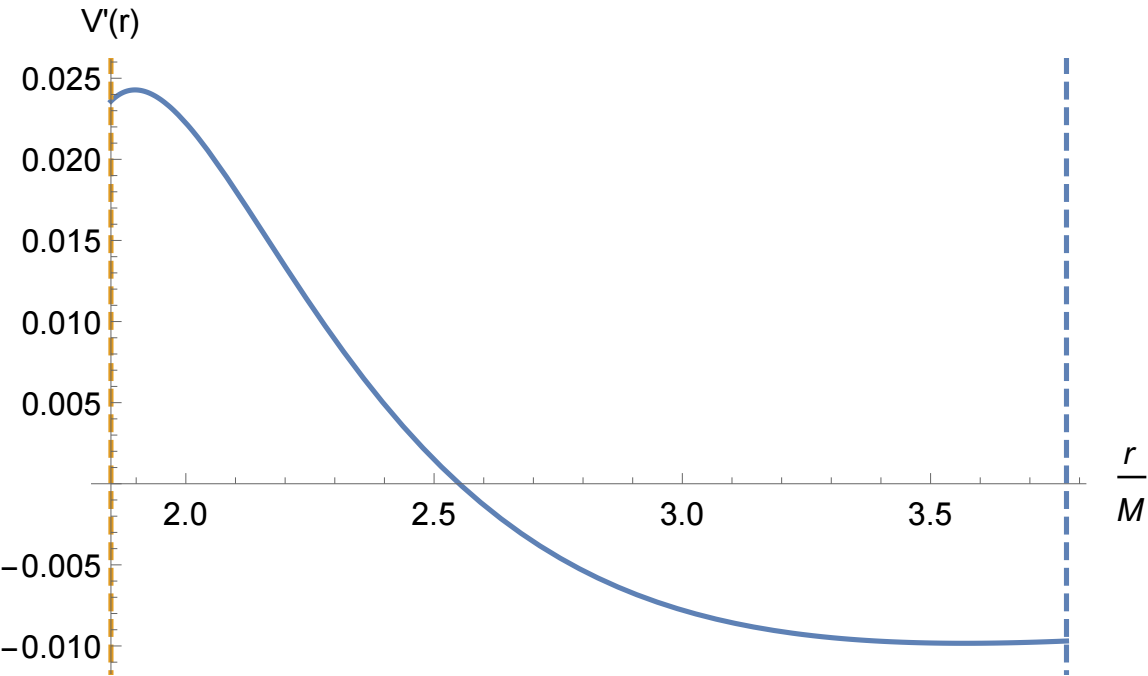}     \includegraphics[width=0.33\textwidth]{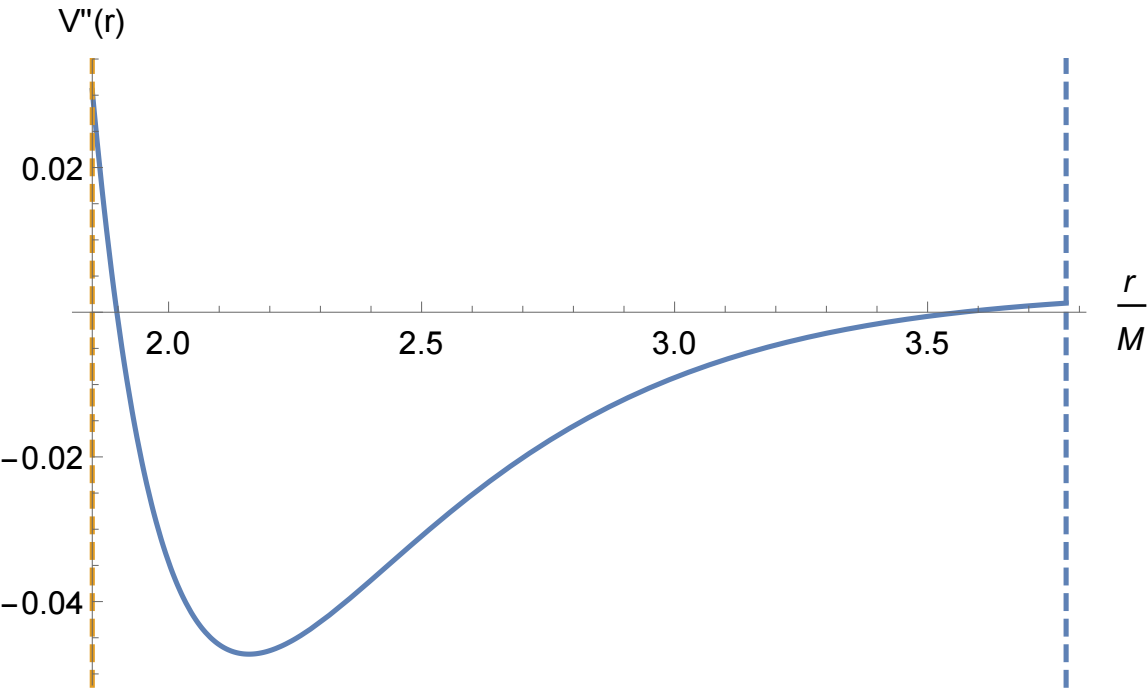}   
\caption{$V(r),V'(r),V''(r)$ as a function of $r$ from $r_{+}$ to $r_{0}$ under parameters $p=0.7,q=0.92$ in S domain.}
        \label{fig:0013}
\end{figure}

\begin{figure}[htbp]
	\centering
\includegraphics[width=0.33\textwidth]{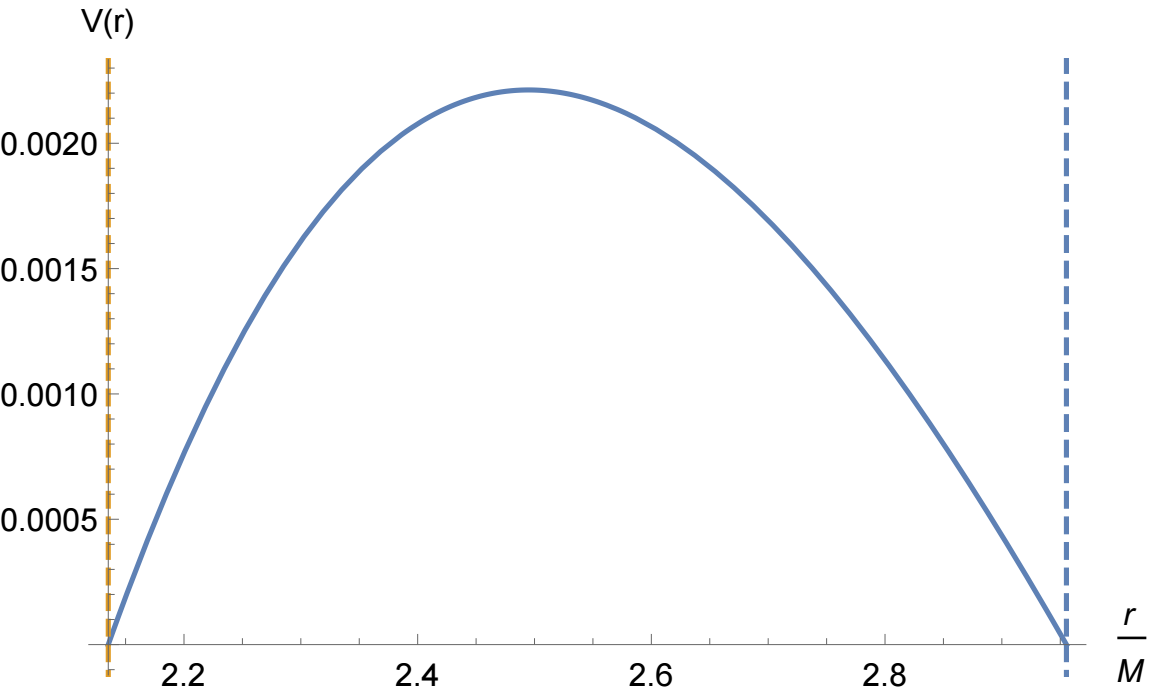}   \includegraphics[width=0.33\textwidth]{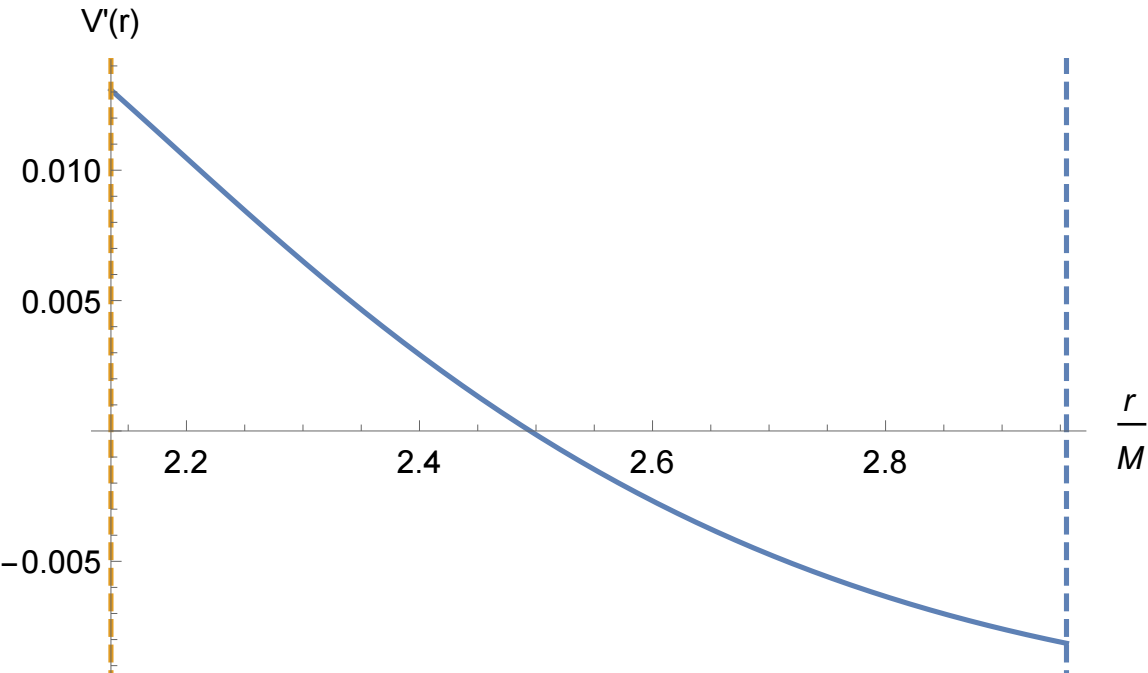}     \includegraphics[width=0.33\textwidth]{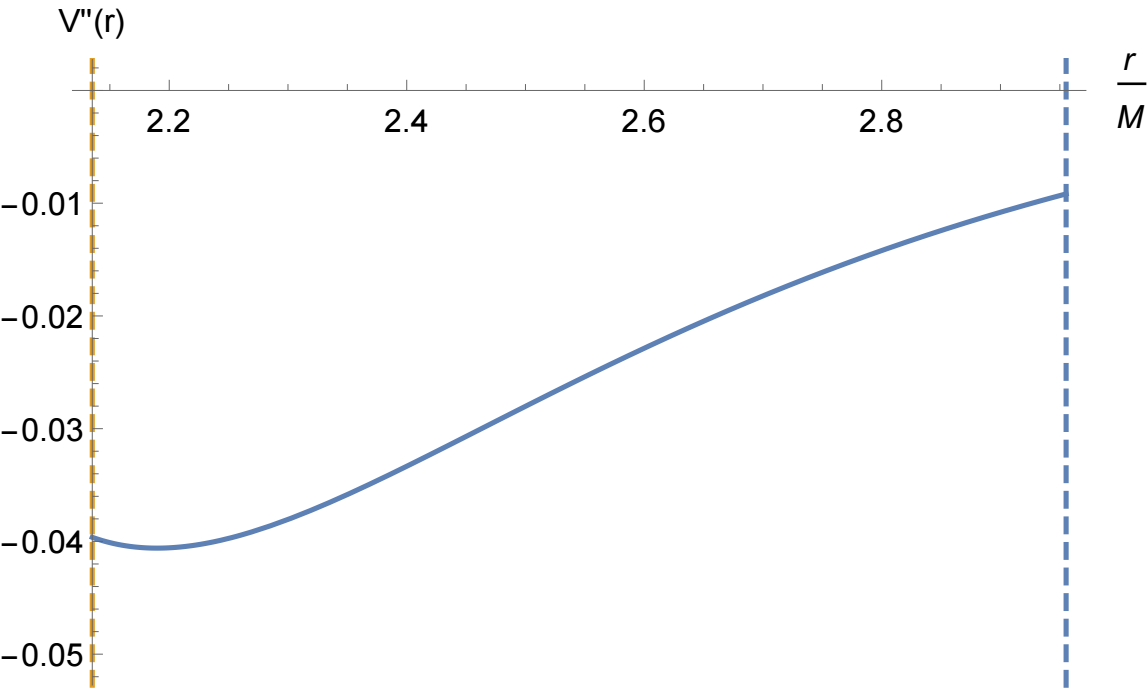}   
\caption{$V(r),V'(r),V''(r)$ as a function of $r$ from $r_{+}$ to $r_{0}$ under parameters $p=q=0.85$ in C domain.}
        \label{fig:0014}
\end{figure}

\begin{figure}[htbp]
	\centering
\includegraphics[width=0.33\textwidth]{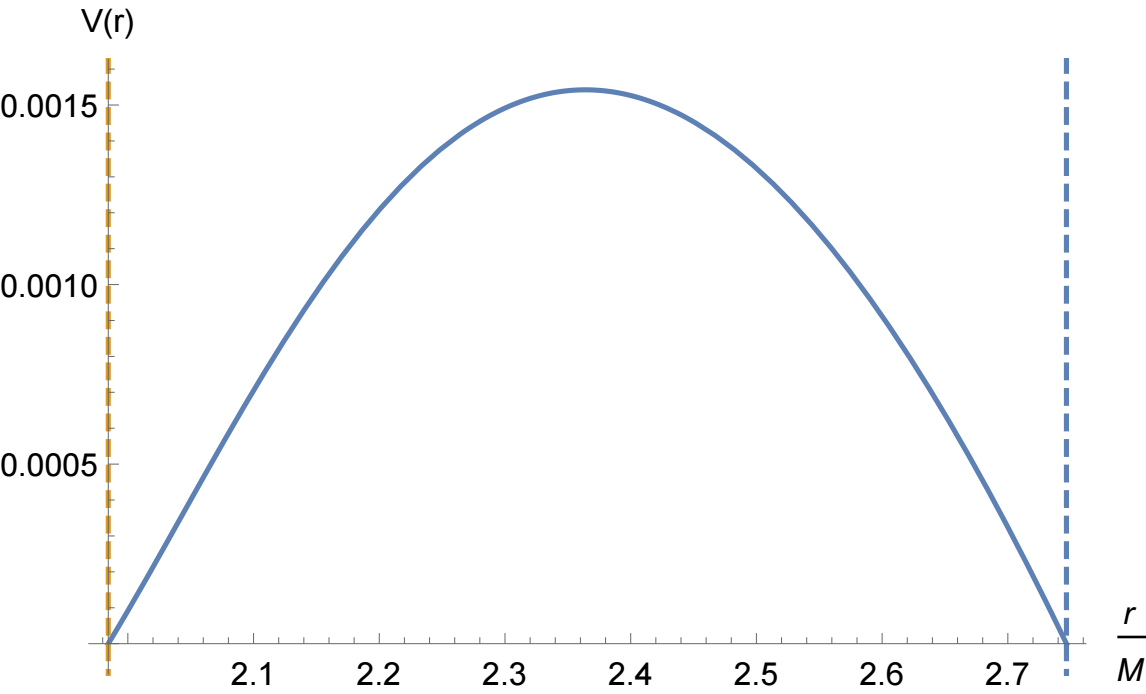}   \includegraphics[width=0.33\textwidth]{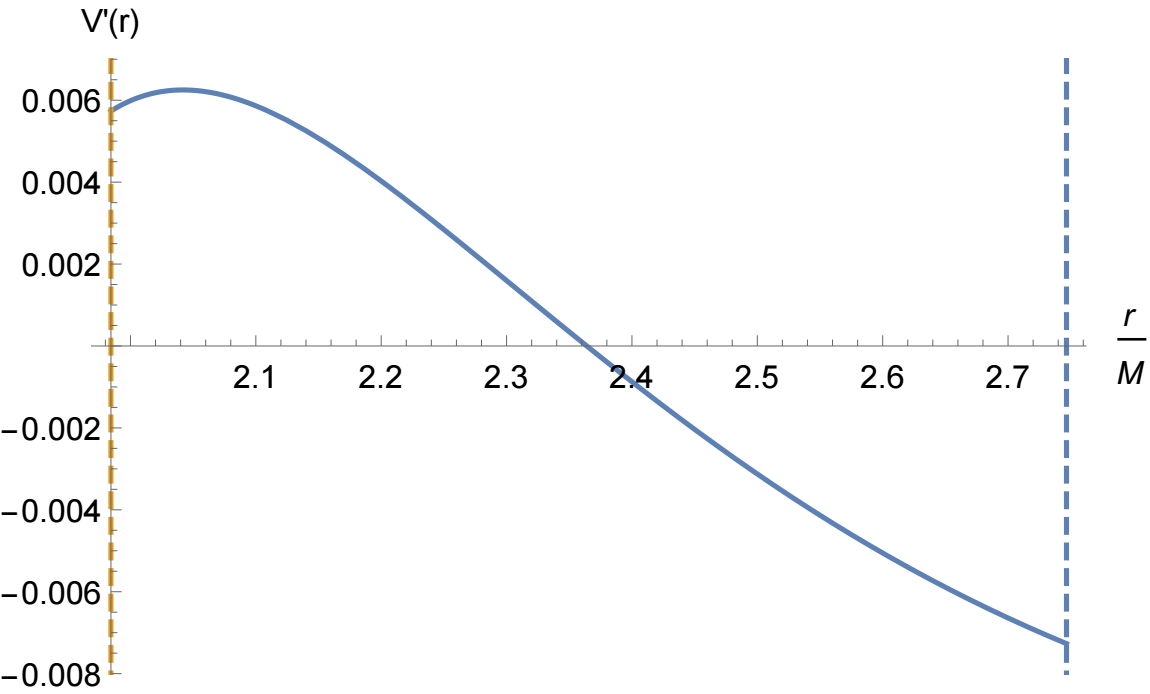}     \includegraphics[width=0.33\textwidth]{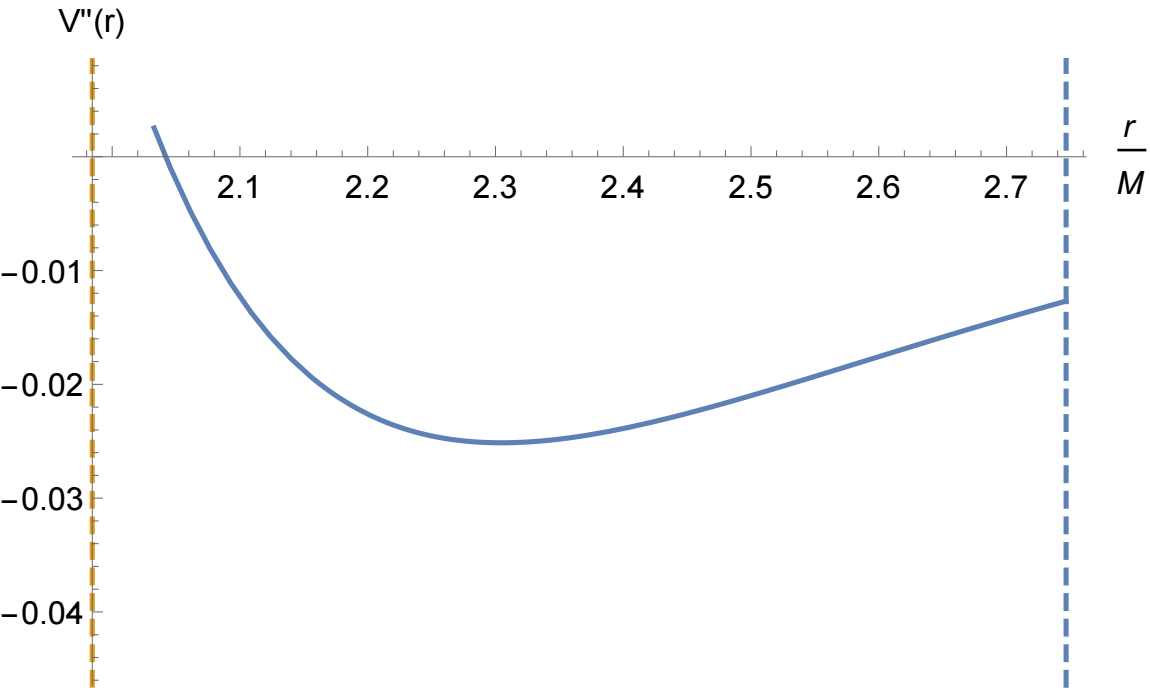}   
\caption{$V(r),V'(r),V''(r)$ as a function of $r$ from $r_{+}$ to $r_{0}$ under parameters $p=0.85,q=0.95$ in D domain.}
        \label{fig:0015}
\end{figure}

For null geodesics, Eq.(\ref{24}) further reduces to
\begin{eqnarray}\label{234}
     (\frac{\mathrm{d}u}{\mathrm{d}\theta})^2=(\frac{1}{D^2}+\frac{\Lambda}{3}-\frac{\alpha\Lambda^2}{36})-u^2+(2M-\frac{\alpha\Lambda M}{3})u^3-\alpha M^2u^6\,,
\end{eqnarray}
and through numeral plotting it's rather simple to confirm that for any value of $p^2,q^2$ in the domain, there exists the only $r^{*}$ satisfying $V'(r^{*})=0$ and $p^2<r^{*}/r_{0}<1$, together with a special impact parameter of $D_{*}=r_{*}/\sqrt{f(r_{*})}$, which means there is one photon ring between the outer horizon and cosmological horizon. For more detail we calculate $V''(r)$ and $V'''(r)$ under different $p$ and $q$. As shown in Fig.\ref{fig:0001} and Fig.\ref{fig:0011}-Fig.\ref{fig:0015}, we actually get six main different kinds of effective potential: only for parameters in region $\mathrm{S}$, $V''(r)$ has two roots between $r_{+}$ and $r_{0}$, while when in region $\mathrm{A,B,C,D}$, $V''(r)$ has one peak but no more than one root and in region $\mathrm{O}$, $V''(r)$ has no peak.

Near the critical radius the asymptotic behavior is
\begin{eqnarray}
  u-u_{*} \rightarrow \mathrm{exp}\Big{(}{-\sqrt{-1+3(2M-\frac{\alpha M\Lambda}{3})u_{*}-15\alpha M^2u_{*}^4}\theta} \Big{)}\,,   
\end{eqnarray}
With a impact parameter smaller than $D_{*}$ there is no $u$ satisfying $\mathrm{d}u/\mathrm{d}\theta=0$, which means there is no perihelion. What's more, for an observer closer to the cosmological horizon, the moving curve of the light rays is more similar to the case of the one in the asymptotic infinity in no cosmological constant case, which can be illustrated by Eq.(\ref{24}), and this is very important in the following light tracing method in drawing the black hole shadow. After transposing in the observer's region (between the outer and cosmological horizon) , the light rays can still approach infinity as in Eq.(\ref{234}) the constant term is always positive in qOS-dS spacetime. 

\begin{figure}[htbp]
	\centering
\includegraphics[width=0.5\textwidth]{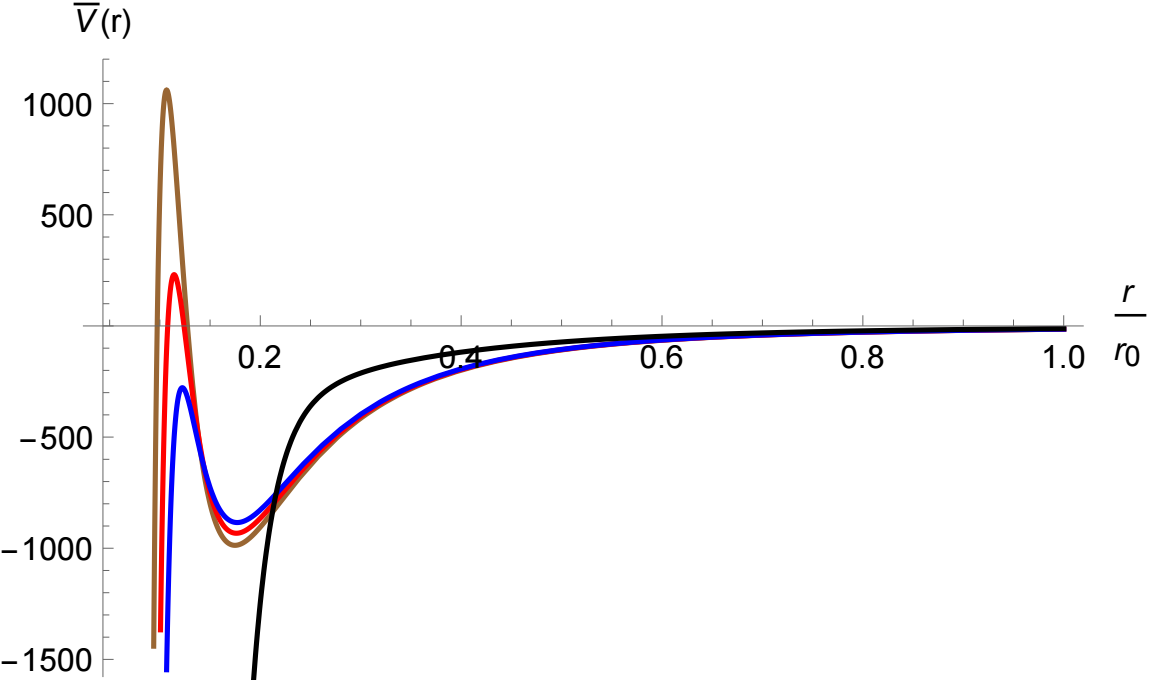}
         
\caption{The relation between the effective potential $\overline{V}$ (the numerator of Eq.(\ref{1090})) and radius $r$ under different parameter (brown for $p=q=0.17$, red for $p=q=0.175$, blue for $p=q=0.18$, black for $p=q=0.3$), and where $\overline{V}$ is positive the circular orbit is stable. The roots of $\overline{V}$ are ISCO and OSCO. }
        \label{fig:0}
\end{figure}

However, for the problem of confirming the ISCO and OSCO of the time-like geodesics things get rather complicated.  We then follow the methods in~\cite{Cao:2024kht}, that is, by solving the equation $V'=\mathrm{d}V(r)/\mathrm{d}r=0$, we can get a function between $L^2$ and $r$, and by drawing its graph one can easily know the radius of a circular orbit must be larger than $r_{0}$. As for the position for ISCO and OSCO, for points on the curve there is always
\begin{eqnarray}
    0=\frac{\mathrm{d}V'(r,L)}{\mathrm{d}t}=V''\frac{\mathrm{d}r}{\mathrm{d}t}+\frac{\partial V'}{\partial L}\frac{\mathrm{d}L}{\mathrm{d}t}\,,
\end{eqnarray}
in this way there is 
\begin{eqnarray}\label{1090}
    V''(r)=-\frac{\partial V'}{\partial L}\frac{\mathrm{d}L}{\mathrm{d}r}=(\frac{2f}{r^3}-\frac{f'}{r^2})\frac{\mathrm{d}L^2}{\mathrm{d}r}
    \end{eqnarray}
    \begin{eqnarray}
    =\frac{2\tilde{c}[-8r^{10}+15\tilde{b}\tilde{c}r_{0}r^9+\tilde{b}r_{0}^3r^7-3(16\tilde{a}+\tilde{b^2})\tilde{c}r_{0}^4r^6+8\Tilde{a}r_{0}^6r^4+9\tilde{a}\tilde{b}\tilde{c}r_{0}^7r^3-24\tilde{a}^2\tilde{c}r_{0}^{10}]}{r^6r_{0}^2(2r^4-3\tilde{b}\tilde{c} r_{0}r^3+6\tilde{a}\tilde{c}r_{0}^4)}\,,
\end{eqnarray}
where $t$ is the parameter of the $r-L$ curve, and parameters $\tilde{a},\tilde{b},\tilde{c}$ are defined by Eq.(\ref{1}), Eq.(\ref{2}) and Eq.(\ref{3}). $V''=0$ is just the point of the ISCO (for smaller $r$) and OSCO (for bigger $r$). However, as shown in Fig.\ref{fig:0}, there are three circumstances: both ISCO and OSCO exist; the two orbits merge in one; no stable orbits. Nevertheless, considering the more realistic case is $\Lambda$ is rather small and the ratio $p^2$ is very close to 0, we can often regard such a spacetime as having both ISCO and OSCO. However, the only case that such a spacetime permits an elliptic orbits is the existence of both ISCO and OSCO, which can give strong restrictions on the possible parameters $\alpha$ and $\Lambda$. 
\begin{figure}[htbp]
	\centering
\includegraphics[width=0.45\textwidth]{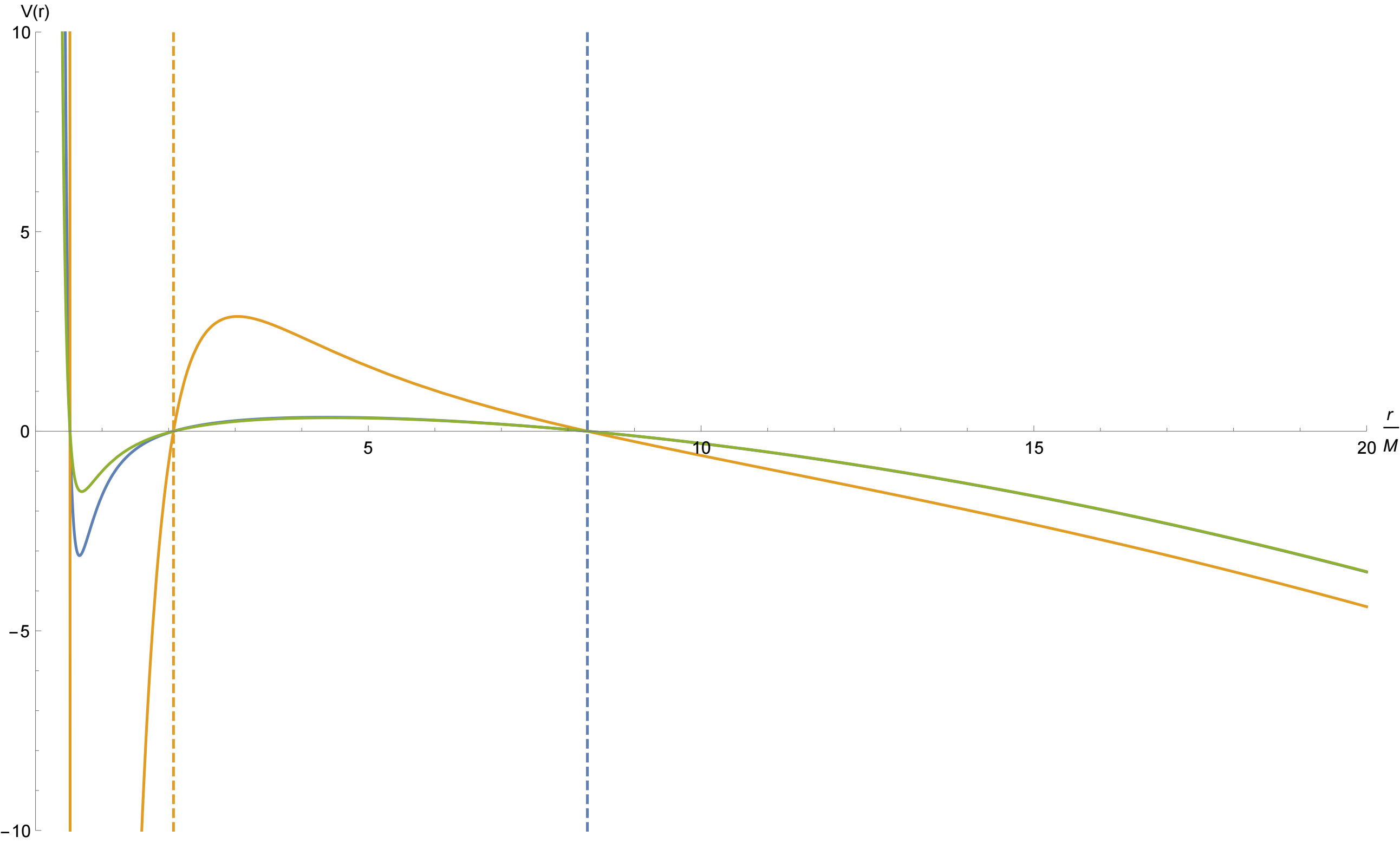}\includegraphics[width=0.45\textwidth]{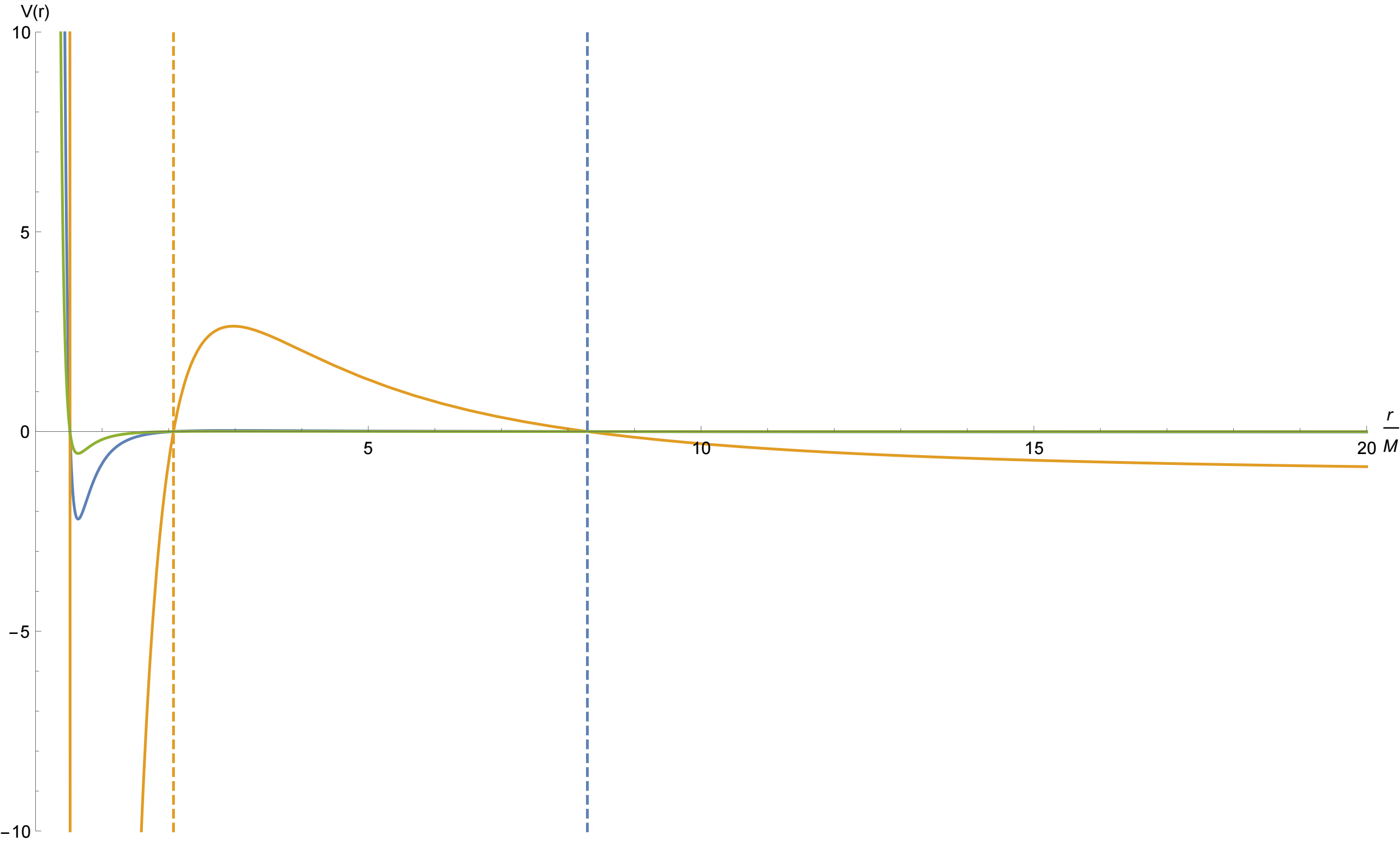}
         
\caption{The relation between the effective potential $V$ and radius $r$ under different parameter $L$ (green for $L=0.5$, blue for $L=1$, yellow for $L=10$) and spacetime parameters $p=q=0.5$, while the left is for time-like geodesics and the right is for null geodesics. The two dashed vertical line mark the position of outer horizon and cosmological horizon.}
        \label{fig:00}
\end{figure}
Another notable fact about the time-like geodesics is that the difference between it and a null geodesic with the same impact parameter when approaching the cosmological horizon disappears. About the asymptotic behavior, just as in the null case the particle can never arrive at $u\rightarrow{\infty}$ (or, equivalently, $r\rightarrow 0$), but for the other side, $u\rightarrow 0$, things are quite different from an asymptotic flat spacetime. From Eq.(\ref{24}) it's easy to find there is 
\begin{eqnarray}
    u\rightarrow -\sqrt{2(\theta-\theta_{0})\sqrt{\frac{1}{L^2}(\frac{\Lambda}{3}-\frac{\alpha\Lambda^2}{36})}}+u_{0}\,,
\end{eqnarray}
near $u\approx 0$, in which $\theta_{0}$ and $u_{0}$ are two constants along the propagation.

\section{generalization to rotational case: geodesics}\label{sec:3}
Though Newman-Janis algorithm, we can easily get the line element of the rotating qOS-dS solution is (one can find detailed derivation in Appendix.\ref{app_2}):
 
\begin{eqnarray}\label{final}
    \mathrm{d}s^2=-(1-\frac{G(r)}{\rho^2})\mathrm{d}t^2+\frac{\rho^2}{\Delta}\mathrm{d}r^2+\rho^2\mathrm{d}\theta^2-2a\mathrm{sin}^2\theta\frac{G(r)}{\rho^2}\mathrm{d}t\mathrm{d}\varphi+\mathrm{sin}^2\theta[\frac{(r^2+a^2)^2-a^2\Delta\mathrm{sin}^2\theta}{\rho^2}]\mathrm{d}\varphi^2\,,
\end{eqnarray}
where
\begin{eqnarray}
    \Delta=r^2f(r)+a^2,\qquad G(r)=r^2(1-f(r)),\qquad\rho^2=r^2+a^2\mathrm{cos}^2\theta\,,
\end{eqnarray}

We start our study on the moving curves from the fact that such a metric in Eq.(\ref{final}) is describing a Petrov-D spacetime, so we can also use the Walker-Penrose theorem to  find constants~\cite{Chad1985}. We can choose Newman-Penrose tetrads as follows (where  $\overline{\rho}= r+\mathrm{i}a\mathrm{cos}\theta$)
\begin{eqnarray}
    l^a=\frac{1}{\Delta}(r^2+a^2,\Delta,0,a)\,,\qquad
    n^a=\frac{1}{2\rho^2}(r^2+a^2,-\Delta,0,a)\,,
\end{eqnarray}
\begin{eqnarray}
    m^a=\frac{1}{\sqrt{2}\overline{\rho}}(\mathrm{i}a\mathrm{sin}\theta,0,1,\mathrm{icsc}\theta)\,,
\end{eqnarray}
then the spin coefficient $\kappa=\sigma=\lambda=\nu=\epsilon=0$ , which, according to Goldberg-Sachs theorem, indicates that just like Kerr spacetime, rotating qOS-dS spacetime is a Petrov-D spacetime. In this case the only non-zero Weyl scalar is $\Psi_{2}=R_{abcd}l^an^bm^c\overline{m}^d$. and we can find scalar function $W$ to satisfy
\begin{eqnarray}
    DW=D|\Psi_{2}|^{-2/3}, \qquad \Delta W=\Delta|\Psi_{2}|^{-2/3},\qquad
    \delta W=\delta^* W=0\,.
\end{eqnarray}
in which $D,\Delta, \delta, \delta^{*}$ are all derivative operators, respectively related to tetrads $l, n, m, \bar{m}$. So again according to Walker-Penrose theorem, no matter for null-geodesics or time-like geodesics with its tangent vector $k^a$ there is a constant along the geodesic called Carter Constant
\begin{eqnarray}
    K=2|\Psi_{2}|^{-2/3}(k^al_{a})(k^{a}n_{a})-W|k|^2\,.
\end{eqnarray}
Another two moving integrals are 
\begin{eqnarray}\label{b}
    E=(1-\frac{G(r)}{\rho^2})\Dot{t}+a\mathrm{sin}^2\theta\frac{G(r)}{\rho^2}\Dot{\varphi}\,,
\qquad
    L=-a\mathrm{sin}^2\theta\frac{G(r)}{\rho^2}\Dot{t}+\mathrm{sin}^2\theta[
    \frac{(r^2+a^2)^2-a^2\Delta\mathrm{sin}^2\theta}{\rho^2}]\Dot{\varphi}\,,
\end{eqnarray}
from Eq.(\ref{b}) we can straightly get 
\begin{eqnarray}\label{b1}
    \Dot{t}=\frac{r^2+a^2}{\rho^2\Delta}[E(r^2+a^2)-aL]-\frac{a}{\rho^2}(aE\mathrm{sin}^2\theta-L)\,,
\qquad
    \Dot{\varphi}=\frac{a}{\rho^2\Delta}[E(r^2+a^2)-aL]-\frac{1}{\rho^2}(aE-\frac{L}{\mathrm{sin}^2\theta})\,,
\end{eqnarray}
and the Hamiltonian satisfies (where \quad$\delta=1$ \quad for time-like and  \quad$\delta=0$ \quad for null) 
\begin{eqnarray}\label{d}
    -2H=\delta=E\Dot{t}-L\Dot{\varphi}-\rho^2\Dot{\theta}^2-\frac{\rho^2}{\Delta}\Dot{r}^2\,.
\end{eqnarray}

To find the explicit form of Carter Constant, we can write Hamilton-Jaccobi equation 
\begin{eqnarray}
    \frac{\partial S}{\partial \tau}+H(x,\frac{\partial S}{\partial x})=
     \frac{\partial S}{\partial \tau}+\frac{1}{2}g^{ij}\frac{\partial S}{\partial x^{i}}\frac{\partial S}{\partial x^{j}}=0\,,
\end{eqnarray}
and suppose 
\begin{eqnarray}
    S=\frac{1}{2}\delta \tau-Et-L\varphi+\mathcal{R}(r)+\Theta(\theta)\,,
\end{eqnarray}
and we can find there is Carter constant
\begin{eqnarray}\label{d}
    \mathcal{K}=(\frac{\mathrm{d}\Theta}{\mathrm{d} \theta})^2+\mathrm{cos}^2\theta(L^2\mathrm{csc}^2\theta-a^2E^2+a^2\delta)
    =-\Delta(\frac{\mathrm{d}\mathcal{R}}{\mathrm{d}r})^2+\frac{1}{\Delta}[(r^2+a^2)E-aL]^2-(L-aE)^2-\delta r^2\,,
\end{eqnarray}
notice that this even without the specific form of $\Delta$ ,or $G(r)$, however, if we aim to use Walker-Penrose theorem itself to prove this it will be very hard to calculate all the nonzero components of Riemann tensor.

Now we can study a special case for geodesics: $r$ or $\theta$ is a constant.

Firstly by introducing the well-known parameters
\begin{eqnarray}
    \chi=\frac{L}{E}\,, \qquad
    \eta=\frac{\mathcal{K}}{E^2}\,, \qquad
    \epsilon=\frac{\delta}{E^2}\,,\qquad
    s=\frac{\tau}{E}\,,
\end{eqnarray}
we can get the effective potential for $r$ and $\theta$ moving: (where $\mu$=$\mathrm{cos}\theta$)
\begin{eqnarray}
 \Big{(}\frac{\mathrm{d}r}{\mathrm{d}s}\Big{)}^2+ \frac{V_{r}}{\rho^4}=0\,, \qquad V_{r}=\Delta((a-\chi)^2+\eta+\epsilon r^2)-(r^2+a^2-a\chi)^2\,,
\end{eqnarray}
\begin{eqnarray}\label{t move}
\qquad
    \Big{(}\frac{\mathrm{d}\mu}{\mathrm{d}s}\Big{)}^2+\frac{V_{\theta}}{\rho^4}=0\,, \qquad
    V_{\theta}=a^2(1-\epsilon)\mu^4+(\chi^2+\eta-a^2+a^2\epsilon)\mu^2-\eta\,,
\end{eqnarray}
$\theta$= constant requires $V_{\theta}=0,V_{\theta}'=0$, from which we can get 
\begin{eqnarray}\label{t}
\mathrm{cos}^2\theta=\frac{1}{a}\sqrt{\frac{-\eta}{1-\epsilon}}\,, \qquad \chi^2=(\sqrt{a(1-\epsilon)}-\sqrt{-\eta})^2\,,
\end{eqnarray}
when $\theta\neq\pi/2$, which actually requires $\eta<0, 1-\epsilon<0$, and at the same time there is $V_{\theta}''=-a^2(1-\epsilon)\mathrm{sin}^2\theta\leqslant0$, which indicates that there is no stable $\theta$=constant orbit. When $\theta=\pi/2$, the require is trivial: $\eta=0$, and no restriction on $\chi$.

Similarly for $r=\tilde{r}$= constant we have
\begin{eqnarray}\label{s}
\Delta\eta=\tilde{r}^4-\Delta\epsilon \tilde{r}^2+2\frac{\Delta'-2\tilde{r}f}{a\Delta'}\Delta \tilde{r}^2u+\tilde{r}^3f\frac{2\epsilon\Delta^2+\tilde{r}^2(\tilde{r}\Delta'-4\Delta)}{a^2\Delta'}\,,
\end{eqnarray}
\begin{eqnarray}\label{t}
    u=a-\chi=\frac{\tilde{r}(2\Delta-\tilde{r}\Delta')\pm\Delta\sqrt{4\tilde{r}^2-2\epsilon\Delta'\tilde{r}}}{a\Delta'}\,,
\end{eqnarray}
where $f, \Delta $ and $\Delta'=\mathrm{d}\Delta/ \mathrm{d}r$ are all specific values on the plane $r=\tilde{r}$
and this actually requires $\tilde{r}\geqslant \frac{1}{2}\epsilon\Delta'$. When the particle is supposed to move on the equatorial plane merely, thus $\eta=0$, by inserting Eq.(\ref{t}) into Eq.(\ref{s}) one can get the relation between the angular velocity $\Omega$ and circular orbit $\tilde{r}$:
\begin{eqnarray}\label{209}
    \chi=a-\tilde{r}\frac{\frac{(2\Delta \tilde{r}f-a^2\Delta')(\tilde{r}\Delta'-2\Delta)}{a\Delta\Delta'}-2\Delta\frac{\Delta'-2\tilde{r}f}{\Delta'a}\pm\sqrt{(\frac{(2\Delta \tilde{r}f-a^2\Delta')(\tilde{r}\Delta'-2\Delta)}{a\Delta\Delta'}-2\Delta\frac{\Delta'-2\tilde{r}f}{\Delta'a})^2-2\frac{\tilde{r}}{\Delta}(\tilde{r}\Delta'-2\Delta)(\frac{a^2\Delta'}{2\Delta}-\tilde{r}f)}}{\frac{a^2\Delta'}{\Delta}-2\tilde{r}f}\,,
\end{eqnarray}
\begin{eqnarray}
    \Omega=\frac{aG(\tilde{r})+\tilde{r}^2f(\tilde{r})\chi}{(\tilde{r}^2+a^2)^2-a^2\Delta-a\chi G(\tilde{r})}\,,
\end{eqnarray}

We now focus on $\epsilon=0$ case, and in Eq.(\ref{209}) we must take "+", that is for light rays particularly. In this case the two parameters both have a relatively simple expression:
 \begin{eqnarray}
\chi=\frac{r^2+a^2}{a}-4r\frac{\Delta}{a\Delta'}=\frac{6\tilde{a}\tilde{c}r_{0}^6\tilde{r}^2+r_{0}^2\tilde{r}^5(2\tilde{r}-3\tilde{b}\tilde{c}r_{0})+a^2[2\tilde{a}\tilde{c}r_{0}^6+\tilde{r}^3(\tilde{b}\tilde{c} r_{0}^3+2r_{0}^2\tilde{r}+4\tilde{c}\tilde{r}^3)]}{a[2\tilde{a}\tilde{c}r_{0}^6+\tilde{r}^3(\tilde{b}\tilde{c}r_{0}^3-2 r_{0}^2\tilde{r}+4\tilde{c}\tilde{r}^3)]}
\end{eqnarray}
\begin{eqnarray}\label{lsf}
    \eta=\frac{\tilde{r}^4}{a^2}(\frac{8\Delta \tilde{r}f'}{\Delta'^2}-1)=-r_{0}^2\tilde{r}^4\frac{36\tilde{a}^2\tilde{c}^2r_{0}^{10}+4\tilde{a}\tilde{c}r_{0}^6\tilde{r}^2(8a^2-9\tilde{b}\tilde{c}r_{0}\tilde{r}+6\tilde{r}^2)+\tilde{r}^5[r_{0}^2\tilde{r}(3\tilde{b}\tilde{c}r_{0}-2\tilde{r})^2-8a^2\tilde{c}(\tilde{b}r_{0}^3-2\tilde{r}^3)]}{a^2[2\tilde{a}\tilde{c}r_{0}^6+\tilde{r}^3(\tilde{b}\tilde{c}r_{0}^3-2r_{0}^2\tilde{r}+4\tilde{c}\tilde{r}^3)]^2}
\end{eqnarray}

\begin{figure}[htbp]
	\centering
\includegraphics[width=0.5\textwidth]{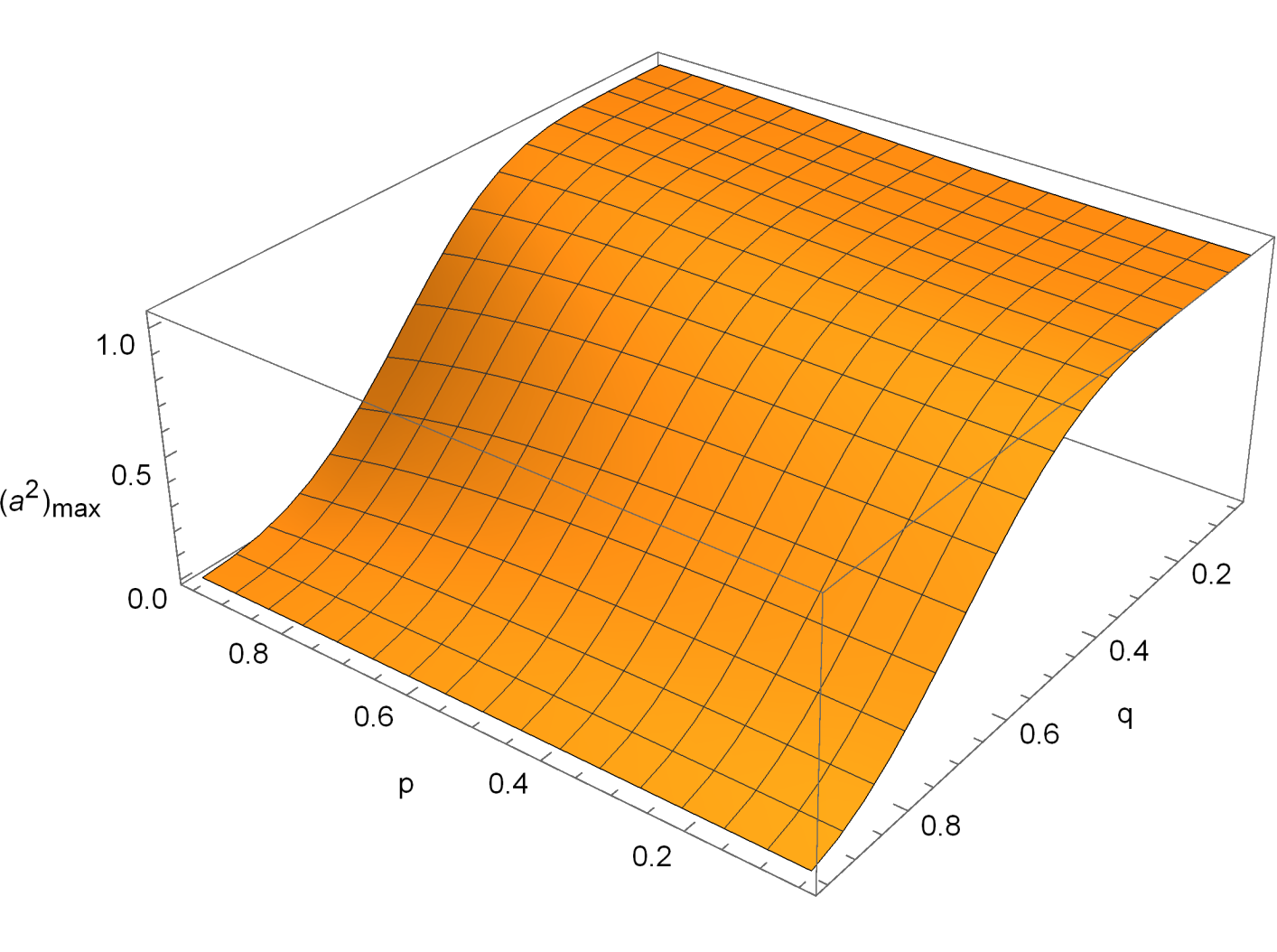}   
\caption{The maximum value of $a^2$ with certain $p$ and $q$.}
        \label{fig:0065}
\end{figure}

Now we need to confirm the range of parameter $a$ under certain parameters $p$ and $q$. If we want to get light on photon ring, considering the accretion disk lies on the equatorial plane ($\theta=0$), it is important to note that among all null circular orbits (whose $\tilde{r}$ can take any positive value in principle) only null geodesics intersecting with the equatorial plane can be actual light rays originated near the black hole, which requires $\eta>0$ for some $\tilde{r}$ according to Eq.(\ref{t move}). Through numerical method it is easy to certify it is true. So there must be $\tilde{r}_{-}<\tilde{r}<\tilde{r}_{+}$, where $\tilde{r}_{-}$ and $\tilde{r}_{+}$ are two bigger root of the function $\eta$. An important  comparison is whether  $\tilde{r}_{\pm}$ are between the outer horizon $r_{+}$ and cosmological horizon $r_{0}$, which are the two roots of function $\Delta$, as all those light originated from other part of the spacetime cannot be observed. From the first part pf Eq.(\ref{lsf}), which indicates that the roots of $\eta$ are inside the region of $\Delta>0$, one is easy to find condition $r_{+}<\tilde{r}_{-}<\tilde{r}_{+}<r_{0}$ is always satisfied. So at this time we only need to guarantee there are three horizons in a certain circumstance, which is shown in Fig.\ref{fig:0065}.  

\section{black hole shadow of rotating qos-ds black hole}\label{sec:4}
\subsection{Fundamental Methods\label{sec:5}}
Here we firstly illustrate the common method used in plotting black hole image when the spacetime is asymptotically flat. According to~\cite{Cao:2024kht}, for time-like geodesics commonly there is ISCO in various spacetime. As for the null geodesics, in spherically symmetric case there are only one or two circular orbits. All these conclusions can be naturally generalized to our rotating black hole, except for the fact that here the image has two parameters observed and the photon ring is actually not corresponding to a sole $r$=constant orbit. About the black hole shadow, in the asymptotic flat spacetime the well-known method of light tracing is always begun with the definition of ``celestial coordinate":
\begin{eqnarray}\label{1119}
    \alpha=-\chi \mathrm{csc}\theta_{0}\,, \qquad
    \beta=\pm \sqrt{\eta+a^2\mathrm{cos}^2\theta_{0}-\chi^2\mathrm{cot}^2\theta_{0}}\,.
\end{eqnarray}

However, in asymptotic dS spacetime things are different, because as the observer position $(r_{I},\theta_{I})$ approaches the cosmological horizon $r_{0}$, the function $\Delta$ approaches to 0 and the following result is expected:
\begin{eqnarray}
    \alpha\rightarrow -\Bigg{(}\frac{r\rho^2\sqrt{\Delta}}{\Sigma^2\mathrm{sin}\theta}\frac{\chi}{1-\omega\chi}\Bigg{)}_{(r_{I},\theta_{I})}\,,
\end{eqnarray}
\begin{eqnarray}
    \beta\rightarrow \pm \Bigg{(}\frac{r\sqrt{\Delta}}{\Sigma}\frac{\sqrt{\eta+a^2\mathrm{cos}^2\theta_{0}-\chi^2\mathrm{cot}^2\theta_{0}}}{1-\omega\chi}\Bigg{)}_{(r_{I},\theta_{I})}\,,
\end{eqnarray}
here $\Sigma^2=(r^2+a^2)^2-a^2\Delta\mathrm{sin}^2\theta$ and $\omega=2a^2r^2G(r)/\Sigma^2$. The ``$\rightarrow$" indicates that the approximation is appropriate when the observer is near the cosmological radius while this radius is so large that both $V_{\theta}/\rho^2$ and $L/r_{0}^2\rightarrow 0$, which leads to a very small relative distortion of nearby light rays, or in the same words light almost transfer in straight lines. Now it's clear that if the observer approaches the cosmological horizon both $\alpha$ and $\beta$ approaches 0. However, no image distortion is expected, as the relative ratio of $\alpha$ and $\beta$ is the same. Given that $\beta$ is proportional to the final derivative of $\theta$, the sign of it indicates different photon orbits. And this leads to the fact that although the critical line of the photon ring on the $\alpha-\beta$ plane is symmetric about the $\alpha$ axis, the final observed image is not. 

Luckily, we still have another set of well-defined parameters, which is presented as follows. Considering now the observer is not placed at spatial infinity as before, we must distinguish different kinds of observers when applying specific calculations. Although conventionally people chose static observer, it is actually more physical to regard a ZAMO observer. The first reason for this lies on the fact that an observer going through a geodesic in a spacetime with killling vector $\phi^{a}$ and its four-velocity $u^{a}$ must satisfy: $\mathcal{L}_{u}(u_{a}\phi^{a})=0\,$ which explicitly gives a moving integral along the geodesic, i.e., the angular momentum. Now ZAMO observers are chosen to satisfy $u_{a}\phi^{a}=0$, which means these observers are real ``no-rotation" observers (this inner product is a geometrical sum and invariant for any diffeomophism), and being relatively static with the black hole guarantees such observers feel no spatial rotation among them as well. According to Einstein's equivalent principle, any physical rules are supposed to be just like in flat spacetime at least at the origin of a Fermi normal coordinate. So when we consider the influence of redshift phenomenon on optical intensity, it's better to study in a normal basis in which there is boost of tetrads of course but no Fermi-Walker movement (spatial rotation).  

As we have claimed in the introduction, the precise result of the image observed is in ZAMO's observer tetrads ${(e_{(0)},e_{(1)},e_{(2)},e_{(3)})}$ or equivalently ${(e_{(t)},e_{(r)},e_{(\theta)},e_{(\varphi)})}$~\cite{PhysRevD.106.064058}, in which we can define four-momentum as
\begin{eqnarray}
p_{(\mu)}=k_{\nu}e_{(\mu)}^{\nu}\,,
\end{eqnarray}
where $k_{\nu}$ is the four-wave-vector of the light. On the other hand, in the tetrads of ZAMO we can define coordinates $\Theta$ and $\Psi$ as the field of view to label each light ray from the disk and finally received by an observer as
\begin{eqnarray}
\mathrm{cos}\Theta=\frac{p_{(1)}}{p_{(0)}}\,,\qquad
\mathrm{tan}\Psi=\frac{p
_{(3)}}{p_{(2)}}\,,
\end{eqnarray}
then the dimensionless observed projection distance is given by
\begin{eqnarray}
    x=-2\mathrm{tan}\frac{\Theta}{2}\mathrm{sin}\Psi\,,
    \qquad
    y=-2\mathrm{tan}\frac{\Theta}{2}\mathrm{cos}\Psi\,.
\end{eqnarray}

\begin{figure}[htbp]
	\centering
\includegraphics[width=0.7\textwidth]{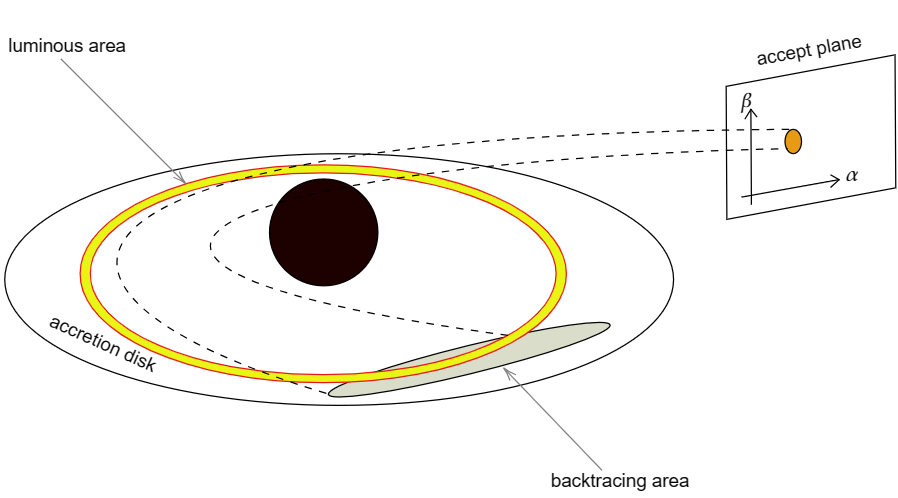}
         
\caption{Schematic diagram for the light tracing method in Kerr-type spacetime.}
        \label{fig:131}
\end{figure}

We follow similar methods in ~\cite{demartino2023optical} to plot an image of the light intensity around the black hole. For simplicity, we assume the disk to be optically thin on the equatorial plane, and its medium have negligible refraction effect. The evolution of the intensity along a light ray is determined by the following equation~\cite{PhysRevD.106.064058}:
\begin{eqnarray}
    \frac{\mathrm{d}(I_{\lambda }/\nu^{3})}{\mathrm{d}\lambda} =\frac{J_{\nu }-\kappa I_{\lambda }}{\nu ^{3}} \,,
\end{eqnarray}
where $\lambda$
 is the affine parameter of null geodesics and $I_{\lambda},  J_{\nu},\kappa_{\nu}$
are the specific intensity, emissivity  and absorption coefficient at the frequency $\nu$, respectively. When the light propagates in vacuum, or they pass through an optically thin disk, both $J_{\nu}$ and $\kappa_{\nu}$ are 0, thus $\frac{I_{\nu}}{\nu^{3}}$ is conserved along the geodesics. Thus in order to study the specific distribution of intensity of the image, firstly we need to study the redshift effect of the light in an arbitrary track near the black hole. To get a physically appropriate result, we set a local Lorenz coordinate of ZAMO observer instead of steady observer~\cite{PhysRevD.106.064058} and the frequency observed by an observer in it is nothing but the No.0 component of wave vector in the tetrads. It has the relation with the moving integral $E$ as follows 
\begin{eqnarray}\label{12356}
    E=\frac{\rho \sqrt{\Delta}}{\Sigma}\omega+\frac{aG(r)}{\Sigma^2}L\,,
\end{eqnarray}
where $\Sigma^2=(r^2+a^2)^2-a^2\Delta \mathrm{sin}^2\theta$. From Eq.(\ref{12356}) we can immediately get the redshift effect for the light emitting from a certain point ($r,\theta$) to another ($r',\theta'$) :
 \begin{eqnarray}\label{lk}
     \gamma=\frac{\omega'}{\omega}=\frac{\frac{\rho\sqrt{\Delta}}{\Sigma}}{\frac{\rho'\sqrt{\Delta'}}{\Sigma'}+a\Big{(}\frac{G}{\Sigma^2}-\frac{G'}{\Sigma'^2}\Big{)}\chi'}\,,
 \end{eqnarray}
 where $\omega'$ and $\omega$ are respectively the frequency observed and emitted, and $\chi'=L/\omega'$. It is notable that the impact parameter in the redshift expression originated from the rotation of the black hole, which means even for two lights with the same frequency emitted from a certain point to meet at another point, their final frequency may be different. An orbit can be solely confirmed given the final $(\alpha,\beta)$, or equally, $(\chi,\eta)$. And if we fix a certain $\theta=0$ for this orbit (say, searching for $n$ th intersecting position of the light ray and the accretion disk), we can get a determinant $(r,\varphi)$. So these two coordinates, together with $L$, can actually be regarded as functions of $(\alpha,\beta)$.  For very distant observer, there is $E=\omega$, so the obtained intensity will be
\begin{eqnarray}\label{1233}
    I_{o}^{(n)}(r(\alpha,\beta),\theta(\alpha,\beta))= \int I^{(n)}(r,\theta,\omega_{o}) \,\mathrm{d}\omega_{o}=\int \frac{\omega_{o}^3}{\omega_{e}^3}\gamma_{n}I(r_{n},\frac{\pi}{2},\omega_{e})\frac{\mathrm{d}\omega_{o}}{\mathrm{d}\omega_{e}}\mathrm{d}\omega_{e}\,.\qquad
\end{eqnarray}
where $r_{n}$ is the  $r$ coordinate for $n$th intersecting of the light rays and the equatorial plane and $\theta_{n}$ now automatically takes $\pi/2$. $\gamma_{n}$ is a to-be-sought solid angle dispersion factor (supposing isotropic luminous model). For light rays that fall on the accept plane far from the critical circular orbits this factor can generally be set 1. For those very close to the critical line, a proper approximation is to replace this factor with a effective area factor, that is, the ratio of the total luminous area to the backtracking area from a tiny area element, or, according to~\cite{PhysRevD.100.024018}, we are actually calculating an average intensity on a certain angle, given that the whole luminous area centers on a small part of the whole backtracking area with an even intensity while other area has none, as shown in Fig.\ref{fig:131}. This condition is almost always satisfied under our model.

Now we move on to prove that the well-known approximation in which only intersections of first two times are taken into account makes sense. As what we study is those light rays close to circular orbits, we can naturally relate the area element ratio described in last paragraph to the critical asymptotic behavior of $r$ moving. We suppose a light ray near the position $\tilde{r}$ with the constants $\tilde{\chi},\tilde{\eta}$ and other physical sums like $\tilde{\Delta}$ and $\tilde{\rho}$, etc. The light then slightly changes its orbit through 
\begin{eqnarray}
    \tilde{r}'=\tilde{r}(1+\delta r)\,,\qquad
    \eta'=\tilde{\eta}(1+\delta \eta)\,,\qquad
    \chi'=\tilde{\chi}(1+\delta \chi)\,,
\end{eqnarray}
and because the circular orbit requires $V_{r}'(\tilde{r})=0$, $\delta \eta$ and $\delta \chi$ are of the same order of $(\delta r)^2$, and in this way there is 
\begin{eqnarray}
    \delta V_{r}=C_{r}(\delta r)^2-\delta B\,,
\end{eqnarray}
where $\delta B=-C_{\eta}\delta \eta-C_{\chi}\delta \chi$ and 
\begin{eqnarray}
    C_{r}=\frac{\tilde{r}^2}{2}\frac{\partial^2 V_{r}}{\partial r^2}(\tilde{r},\tilde{\chi},\tilde{\eta})\,,\qquad  C_{\chi}=\tilde{\chi}\frac{\partial V_{r}}{\partial \chi}(\tilde{r},\tilde{\chi},\tilde{\eta})\,,\qquad C_{\eta}=\tilde{\eta}\frac{\partial V_{r}}{\partial \eta}(\tilde{r},\tilde{\chi},\tilde{\eta})\,,
\end{eqnarray}
and we only consider the case that $\delta B>0$, i.e., in which the perturbed light rays can still be regarded as having a relatively stable departure from the circular orbit. The turning points are where there is $\delta V_{r}=0$, so the integrating domain of $\delta r$ is $-\sqrt{\delta B/C_{r}}<\delta r<\sqrt{\delta B/C_{r}}$ and as we mainly focus on light rays moving near the equatorial plane the total change in the line element ($\Delta s$) during two adjacent departure is given by  
\begin{eqnarray}
    \Delta s=\int^{\sqrt{\delta B/C_{r}}}_{-\sqrt{\delta B/C_{r}}} \frac{\tilde{\rho}^2\mathrm{d}\delta r}{\sqrt{-\delta V_{r}}}\approx \frac{(\tilde{r}^2+a^2)\tilde{r}}{\sqrt{C_{r}}}\mathrm{log}\Big{(}\frac{4C_{r}(\delta r)^2}{\delta{B}}\Big{)}\,,
\end{eqnarray}
together with change in the longitude angle being 
\begin{eqnarray}\label{00000}
    \Delta \varphi\approx \frac{\tilde{r}^2}{\tilde{\Delta}}(a(1-\tilde{f})+\chi \tilde{f})\frac{\tilde{r}}{\sqrt{C_{r}}}\mathrm{log}\Big{(}\frac{4C_{r}(\delta r)^2}{\delta{B}}\Big{)}\,,
\end{eqnarray}
however for the latitude angle things are rather complicated, as for the movement in this dimension we can no longer regard $\theta$ as a constant as there is no $\Delta \theta \ll\theta$. For exact solutions one can refer to~\cite{Chad1986}. Here we consider the case that $\mu$ ($\mu=\mathrm{cos}\theta$) changes a little sum from $-\Delta\mu/2$ to $\Delta\mu/2$, then through the expansion of the first type of Jacobian function $F(\psi,k)$ (where $\mathrm{cos}\psi$ is proportional to $\mu$ and $k$ is a function of $\chi,\eta$ and $a$) near $\psi=\pi/2$, we get
\begin{eqnarray}\label{11111}
    \Delta \mu\approx\frac{\tilde{\eta}\tilde{r}}{a^2\sqrt{C_{r}}}\mathrm{log}\Big{(}\frac{4C_{r}(\delta r)^2}{\delta{B}}\Big{)}\,.
\end{eqnarray}
For light rays that are very close to moving completely on the equatorial plane (which is the most realistic condition for actual observation), $\eta$ is such a small sum that while $\varphi$ changes in a large amount (see Eq.(\ref{00000})), $\theta$ changes a little (see Eq.(\ref{11111})), so we can trustingly suppose between adjacent intersections with the accretion disk the diversion angle of $\varphi$ is a big sum, for instance, $\pi$. Thus according to the processing method in~\cite{PhysRevD.100.024018}, we have a exponential decay series consisted of contributions from every times of intersection:
\begin{eqnarray}\label{12309}
    I_{total}=I_{local}+I_{disk}(1+e^{-\varphi_{0}}+e^{-2\varphi_{0}}+\cdots)\approx I_{local}+I_{disk}\,,
\end{eqnarray}
in which $I_{local}$ and $I_{disk}$ is respectively the direct observed light rays and light rays from lensed ring. Any higher order term of the in Eq.(\ref{12309}) can be disregarded and the fist two term has no extra intensity factor, i.e., $\gamma_{1}=\gamma_{2}=1$ while $\gamma_{n}=0(n\geqslant3)$.
\subsection{Simulating Results\label{sec:6}}

\begin{figure}[htbp]
\centering
\includegraphics[width=0.35\textwidth]{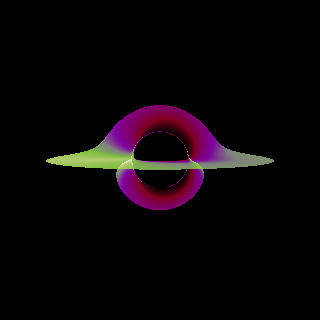}\qquad
\includegraphics[width=0.35\textwidth]{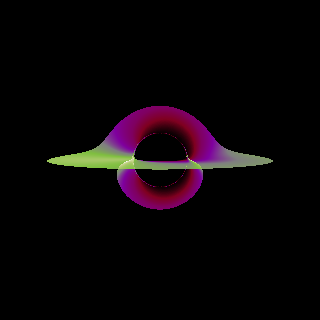}
\includegraphics[width=0.35\textwidth]{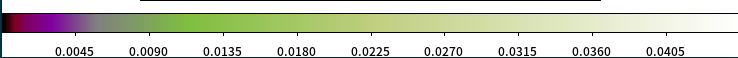}\qquad
\includegraphics[width=0.35\textwidth]{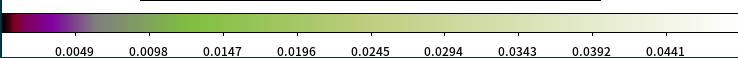}
\includegraphics[width=0.35\textwidth]{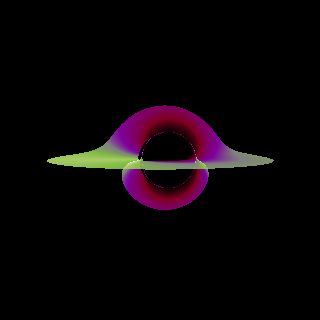}\qquad
\includegraphics[width=0.35\textwidth]{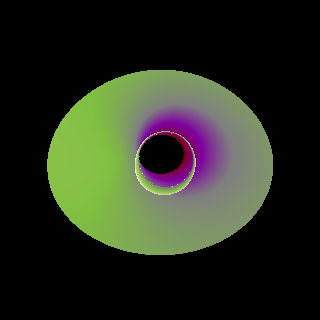}
\includegraphics[width=0.35\textwidth]{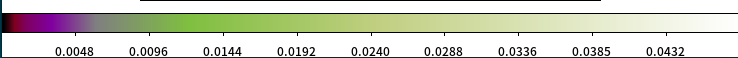}\qquad
\includegraphics[width=0.35\textwidth]{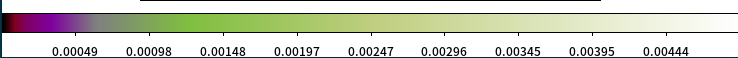}    
\caption{Comparisons of the images of the qOS-dS rotating black holes under different parameters. All these figures are taken at the distance $d=93.91$ with a field of view of $0.628$, while the thin disk range from $r=3$ to $r=20$. The first three panels are observed in the direction $\theta_{0}=1.5$, while the last one is observed at the direction $\theta_{0}=0.7$. The first two corresponds to $q=p=0, a=0.01$ (Kerr) and $p=0.05, q=0.99, a=0.01$ respectively, while the last two both correspond to $p=q=0.05, a=0.8$.}
        \label{fig:bs1}
\end{figure}

\begin{figure}[htbp]
\centering
\includegraphics[width=0.35\textwidth]{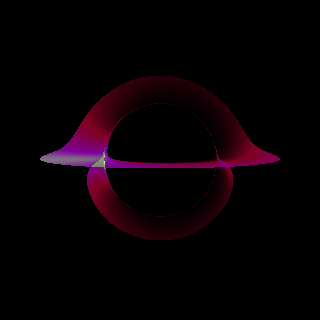}\qquad
\includegraphics[width=0.35\textwidth]{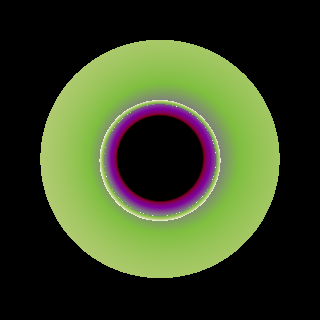}
\includegraphics[width=0.35\textwidth]{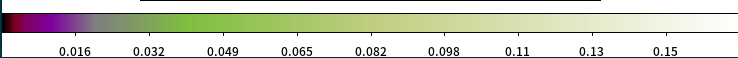}\qquad
\includegraphics[width=0.35\textwidth]{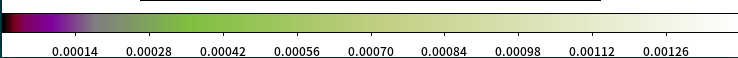}
\includegraphics[width=0.35\textwidth]{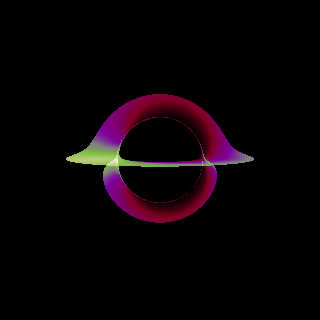}\qquad
\includegraphics[width=0.35\textwidth]{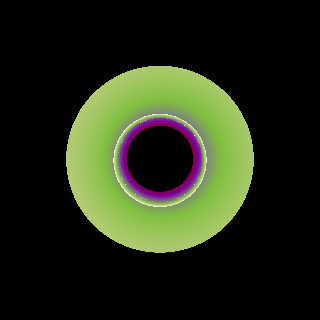}
\includegraphics[width=0.35\textwidth]{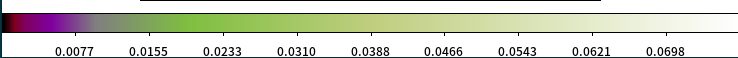}\qquad
\includegraphics[width=0.35\textwidth]{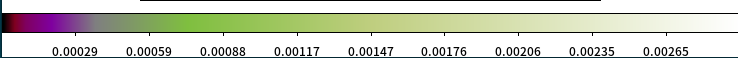}
\includegraphics[width=0.35\textwidth]{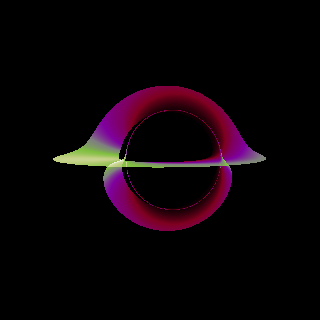}\qquad
\includegraphics[width=0.35\textwidth]{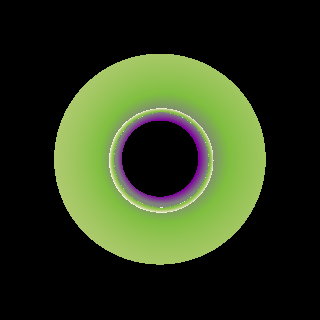}
\includegraphics[width=0.35\textwidth]{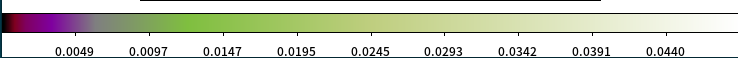}\qquad
\includegraphics[width=0.35\textwidth]{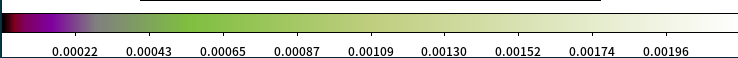}    
\caption{Comparisons of the images of the qOS-dS rotating black holes under different parameters. All these figures are taken at the distance $d=93.91$ with a field of view of $0.314$, while the thin disk range from $r=3$ to $r=10$. The three panels on the right column are all observed at $\theta_{0}=0.1$ while the three panels on the left column are all observed at $\theta_{0}=1.5$. The two panels in the top row corresponds to $q=p=0, a=0.01$ (Kerr), while the two panels in the middle row correspond to $p=0.1, q=0.99, a=0.01$, and the two panels in the bottom row correspond to $p=0.1, q=0.05, a=0.8$.}
        \label{fig:bs2}
\end{figure}

\begin{figure}[htbp]
\centering
\includegraphics[width=0.35\textwidth]{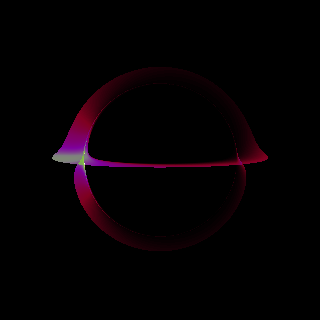}\qquad
\includegraphics[width=0.35\textwidth]{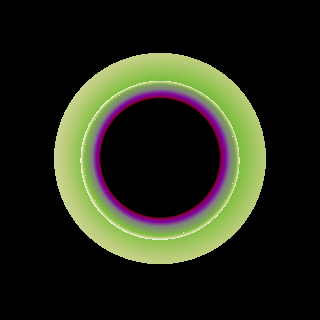}
\includegraphics[width=0.35\textwidth]{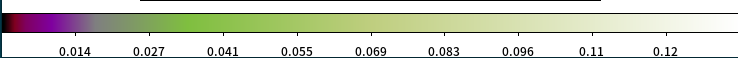}\qquad
\includegraphics[width=0.35\textwidth]{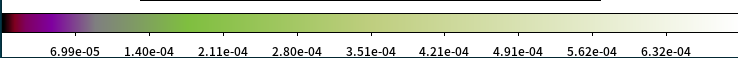}
\includegraphics[width=0.35\textwidth]{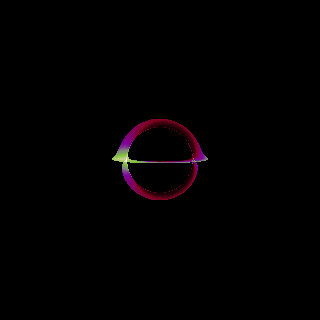}\qquad
\includegraphics[width=0.35\textwidth]{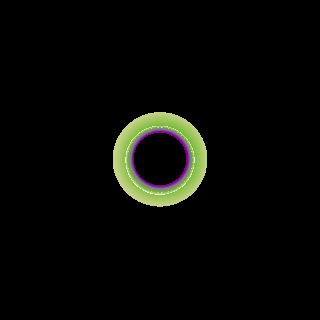}
\includegraphics[width=0.35\textwidth]{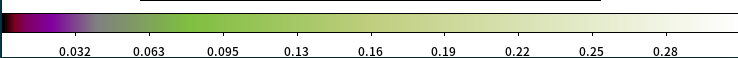}\qquad
\includegraphics[width=0.35\textwidth]{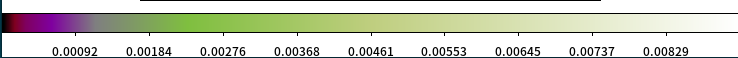}   
\caption{Comparisons of the images of the qOS-dS rotating black holes under different parameters. All these figures are taken at the distance $d=33.91$ with a field of view of $0.628$, while the thin disk range from $r=3$ to $r=6$. The two panels on the right column are both observed at $\theta_{0}=0.1$ while the two panels on the left column are all observed at $\theta_{0}=1.5$. The two panels in the top row correspond to $q=p=0, a=0.01$ (Kerr), while the two panels in the bottom row correspond to $p=0.2, q=0.99, a=0.01$.}
        \label{fig:bs3}
\end{figure}

We use the algorithm GYOTO provided by~\cite{Vincent:2011wz} to plot the image observed by a distant observer. 
This algorithm provides a systematic and effective implement to plot an image of the black hole. Based on an open-access edition on Debian system, people are able to import some structural information about the spacetime and astronomical object and attain a graph as a result. The simulations are presented in \ref{fig:bs1}, \ref{fig:bs2} and \ref{fig:bs3}. The conclusions and discussions are as follows:
\begin{enumerate}
    \item The typical mode of image in de-Sitter spacetime is observed as expected. Apart from the most common ISCO, which restrict the range a thin disk can have, and thus contributes to the distribution and shape of the image accepted, there is now a new boundary of the stable matter circular belt called OSCO, which marks the outer edge of a image. This treatment is physically rational considering the common models (such as in~\cite{demartino2023optical}) describing the accretion disk around the black hole all have a sharp exponential decay as the radiance increases. Yet we are unable to describe the matter's lying even out of the boundary as they are not radially static. In the hands-on work this is particularly prominent as when we set the condition of the thin disk if we permit it to extend to infinity a white edge out of the image is unavoidable.
    \item Through careful comparisons of the first four panels of Fig.\ref{fig:bs2}, there could be seen that a nonzero correction from $p$ can make the light intensity distribution of the whole image become more average. At the same time, there is no surprise that the size of the image becomes significantly smaller when the correction exists. Still, there are no obvious distortion of the shape. Unfortunately, the influence of the parameter $p$ is much more important than that of parameter $q$, while the main effects of the quantum correction is right reflected by $q$. Still, according to the second figure of Fig.\ref{fig:001}, at the near-extrema case (i.e., with $q$ rather large), parameter $p$ also sensitively relies on parameter $\alpha$, even with no much change in cosmological term $\Lambda$. Also, with quantum correction of $q$ the image will be a little smaller (see the top row Fig.\ref{fig:bs1}). Such evidence provide astronomical detectable features in ruling out possible quantum correction with more and more precise precision in the future.
     \item In the comparison of the four panels in Fig.\ref{fig:bs3}, we are able to discover the effect of being closer to cosmological horizon. Although in Fig.\ref{fig:bs2} this might be concealed by other effects, here obviously the light ray intensity of the image becomes much more stronger than Kerr case (note the legends rather than the color of the figures). This is also in accordance to previous results in which an observer near the cosmological horizon is expected to receive a much smaller yet much lighter image, due to the redshift/blueshift effect near the horizon. In this circumstances the redshift factor provided by Eq.(\ref{lk}) is particularly significant. As discussed in the previous paragraph, the quantum correction can have obvious influence of effective cosmological term $p$ even when fixing $\Lambda$. So beyond simply image shape light intensity can also offer potential test on possible correction.
    \item The photon ring, which is due to the critical behavior of light rays near the peak of the potential, becomes more distinguished when the correction exists. This result accords to those which have been proposed about the photon ring in de-Sitter spacetime, say,~\cite{Cao:2024kht}. Also, under the premise that a inner horizon is existent, there might be light rays coming from another universe and transmit through the wormhole area of the spacetime. Yet according to plenty of previous studies, (for this specific kind of metric, see~\cite{PhysRevD.109.064012}), the condition is similar to that of RN, in which a bunch of radiation could end up in infinite mass inflation and cause new singularities near Cauchy horizon, so this spacetime bridge cannot be stable. 
   
\end{enumerate}

\section{conclusions and discussions\label{sec:6}}

In this paper we mainly study the image received by an observer of a certain distance in the spacetime generalized to rotating case using Newman-Janis algorithm from quantum Oppenheimer-Snyder–de Sitter spacetime (qOS-dS). In the first part of our study we present some well-known results about the behavior of null geodesics in spherically symmetric spacetime with this specific metric. The difference of light ray moving from that in asymptotically flat spacetime is noted. Also, a careful classification of different kinds of effective potential with different parameter range is discussed. No matter in what range, there is only one peak between the outer horizon and the the cosmological horizon.

We then generalize the solution from spherically symmetric case to rotating case using Newman-Janis algorithm. We use the common algorithm of analysing the geodesic behavior of a Kerr-type spacetime, and it is shown that most conclusions are alike. We then confirm the physically rational demand for the value of the spin parameter given two other parameters $p$ and $q$, in which not only three horizons exist, but the observer is guaranteed to receive light from equatorial plane as well.

We also discuss specific black hole shadow of qOS-dS rotating spacetime, and through similar analytical calculation methods given by~\cite{PhysRevD.100.024018}, we are able to prove the rationality of merely considering the first few times of a light ray trace passing through the equatorial plane. This is important for the simplification of later computation, and it can shorten the time needed to simulate an image. After all these preparations were done we move on to simulating results, in which we are managed to discover several meaningful alterations of the image from that of Kerr black hole, including their intensity, sizes and shapes.  

There remain a lot of problems to be solved and investigated. First is the strict proof of the rationality of extending the original spacetime to rotating case. Though this has been given by~\cite{PhysRevD.90.064041} in GR framework with the premise that originally there is $g_{00}g_{11}=-1$ in spherically symmetric spacetime, it should be clarified if the same process is correct for LQG framework. Before that we can only know this is approximately rational. Also, such extension needs many premise and is not even effective when considering Kerr-dS spacetime! Second is about the problem of defining and simulating light distortion phenomenon when we regard some light rays transferring through the wormhole from another universe, which might be able to contribute to multi photon rings observed which has been discussed for many times in previous works. Consider the infinite angles a light ray rotate when the spin parameter $a$ is nonzero, then no matter technically or physically there will be new difficulties for drawing the correct image of such a condition. Third, due to the limitations of approximation of light ray tracks in the spacetime, we are still unable to give more nontrivial analytical and quantitative results beside our qualitative conclusions from straight observation.

\section{acknowledgement}
The author is grateful to useful discussions with Long-Yue Li, Yu-Sheng Zhou and others. Also, he thanks for some assists from Cheng-Hao Li, and the useful software provided by F.H.Vincent, T.Paumard, E.Gourgoulhon, and G.Perrin for tracing light rays in a general curved spacetime. 

\appendix
\section{The Derivation of Rotational Proca black hole Solution}\label{app_2}
We shall now generalize our spherical symmetric black hole to rotational black hole based on the Newman-Jains algorithm. However, as remarked in~\cite{PhysRevD.90.064041}, the base of this method critically rely on the fact that in pre-generalized solution there is $g_{00}=1/g_{11}$, as long as we persist on a spherically symmetric premise. For most cases in general Einstein-Proca theory the solution cannot be naturally generalized to Kerr-like Petrov-D spacetime and all elegant conclusions for Kerr black hole may no longer make sense. Lukily for at least the solution in this study, we can still make use of this convenience. Following the standard formalism, first we transform Boyer-Lindquist (BL) coordinates $(t,r,\theta,\phi)$ to  Eddington-Finkelstein (EF) coordinates $(u,r,\theta,\phi)$
\begin{eqnarray}
\mathrm{d}u=\mathrm{d}t-\frac{\mathrm{d}r}{f(r)}\,,
\end{eqnarray}
By introducing Newman-Penrose tetrads 
\begin{eqnarray}\label{ln}
l^a=\Big{(}\frac{\partial}{\partial r}\Big{)}^a\,,
\qquad n^a=\Big{(}\frac{\partial}{\partial u}\Big{)}^a-\frac{f(r)}{2}\Big{(}\frac{\partial}{\partial r}\Big{)}^a\,,\qquad
m^a=\frac{1}{\sqrt{2}|r|}[\Big{(}\frac{\partial}{\partial \theta}\Big{)}^a+\frac{\mathrm{i}}{\mathrm{sin}\theta}\Big{(}\frac{\partial}{\partial \phi}\Big{)}^a]\,,
\end{eqnarray}
we have 
\begin{eqnarray}
g_{ab}=-n_{a}l_{b}-n_{b}l_{a}+m_{a}\overline{m}_{b}+m_{b}\overline{m}_{a}\,.
\end{eqnarray}

By coordinate transformation
\begin{eqnarray}
u'=u-\mathrm{i}a\mathrm{cos}\theta\,,\qquad
r'=r+\mathrm{i}a\mathrm{cos}\theta\,, \qquad
\theta'=\theta\,,\qquad
\phi'=\phi\,,
\end{eqnarray}
we can transform tetrads in (\ref{ln}) to 
\begin{eqnarray}
m^a=\frac{1}{\sqrt{2}|r|}[\mathrm{isin}\theta\Big{(}\frac{\partial}{\partial u}\Big{)}^a-\mathrm{isin}\theta\Big{(}\frac{\partial}{\partial r}\Big{)}^a+\Big{(}\frac{\partial}{\partial \theta}\Big{)}^a+\frac{\mathrm{i}}{\mathrm{sin}\theta}\Big{(}\frac{\partial}{\partial \phi}\Big{)}^a]\,,\qquad
\end{eqnarray}
at the same time we transform $f(r)$ to
\begin{eqnarray}
    \mathcal{F}(r)=\frac{f(r)r^2+a^2\mathrm{cos}^2\theta}{r^2+a^2\mathrm{cos}^2\theta}\,,
\end{eqnarray}
thus we can easily get the inverse matrix of the metric as (where\quad$\rho^2=r^2+a^2\mathrm{cos}^2\theta$)
\begin{eqnarray}
    g^{uu}=\frac{a^2\mathrm{sin}^2\theta}{\rho^2},\quad
    g^{rr}=\mathcal{F}+\frac{a^2\mathrm{sin}^2\theta}{\rho^2},\quad
    g^{ur}=-1-\frac{a^2\mathrm{sin}^2\theta}{\rho^2},\quad
    g^{\theta\theta}=\frac{1}{\rho^2},\quad
    g^{u\phi}=-g^{r\phi}=\frac{a}{\rho^2},\quad
    g^{\phi\phi}=\frac{1}{\rho^2\mathrm{sin}^2\theta}\,.
\end{eqnarray}

In this way the line element becomes
\begin{eqnarray}
\mathrm{d}s^2=-\mathcal{F}\mathrm{d}u^2-2\mathrm{d}u\mathrm{d}r+2a(\mathcal{F}-1)\mathrm{sin}^2\theta\mathrm{d}u\mathrm{d}\phi+2a\mathrm{sin}^2\theta\mathrm{d}r\mathrm{d}\phi+\rho^2\mathrm{d}\theta^2+\mathrm{sin}^2\theta[\rho^2+(2-\mathcal{F})a^2\mathrm{sin}^2\theta]\mathrm{d}\phi^2\,.
\end{eqnarray}
Finally we transform back to BL coordinates by (where \quad$\Delta=r^2f(r)+a^2$)
\begin{eqnarray}
    \mathrm{d}u=\mathrm{d}t-\frac{a^2+r^2}{\Delta}\mathrm{d}r\,,\qquad
    \mathrm{d}\phi=\mathrm{d}\varphi-\frac{a}{\Delta}\mathrm{d}r\,,
\end{eqnarray}
and we can get Eq.(\ref{final}).

\bibliography{reference}{}
\bibliographystyle{apsrev4-1}
\end{document}